\documentclass[journal,final,onecolumn,12pt,twoside]{IEEEtranTCOM}

\usepackage{cite}
\usepackage{graphicx}
\usepackage{subfigure}
\usepackage{amsmath}
\usepackage{amssymb}
\usepackage{bm}
\usepackage{epsfig}
\usepackage{times}
\usepackage{color}
\usepackage{url}
\usepackage{array}
\usepackage{multirow}
\usepackage{rotating}
\usepackage{extarrows}

\usepackage{algpseudocode}
\usepackage[ruled, vlined, linesnumbered]{algorithm2e}
\usepackage{tabularx}
\usepackage{epstopdf}
\usepackage{amsfonts}
\usepackage{lipsum}
\usepackage{mathtools}
\usepackage{cuted}
\usepackage{booktabs, multirow}
\usepackage{siunitx}

\date{}

\usepackage{setspace}
\usepackage{tabulary}
\usepackage{float}

\newtheorem{pavikl}{\textbf{Lemma}}

\newtheorem{pavikp}{\textbf{Proposition}}

\newcommand{\argmin}{\operatornamewithlimits{argmin}}

\newcommand{\halfrac}{\frac{1}{2}}

\renewcommand{\arraystretch}{1.5}

\newcommand{\Rmnum}[1]{\expandafter\@slowromancap\romannumeral #1@}

\begin{document}
\title{D2D Multicast in Underlay Cellular Networks with Exclusion Zones}
\author{\IEEEauthorblockN{Ajay Bhardwaj and Samar Agnihotri}%

\IEEEauthorblockA{School of Computing and Electrical Engineering, IIT Mandi, HP 175$\,$005, India}%

Email: ajay\_singh@students.iitmandi.ac.in, samar.agnihotri@gmail.com%
}
\maketitle

\begin{abstract}
Underlay device-to-device (D2D) multicast communication has potential to improve performance of cellular networks. However, co-channel interference among cellular users (CUs) and D2D multicast groups (MGs) limits the gains of such communication. Allowing the CUs to have exclusion zones around them where no receiver of any MG can exist, is a realistic and pragmatic approach to reduce the co-channel interference of cellular transmission on D2D multicast reception. We use a stochastic geometry based approach to model this scenario. Specifically, we model the locations of CUs and D2D MG receivers with homogeneous Poisson Point Process (PPP), and Poisson Hole Process (PHP), respectively. We formulate the network sum throughput maximization problem in terms of a joint MG channel and power allocation problem with constraints on cellular and MG users’ maximum transmit and acceptable quality of service. We establish that the MG channel allocation problem has computational complexity that is exponential in both, the number of MGs and the number of available cellular channels. Then, we decompose this problem into two subproblems: subset selection problem and subset channel assignment problem. Based on observations and insights obtained from numerical analysis of the optimal solution of the subset selection problem in wide variety of scenarios, we propose a computationally efficient scheme that achieves \textit{almost} optimal performance for the subset selection problem. We further provide a computationally efficient algorithm that achieves \textit{almost} optimal performance for the subset channel assignment problem. Finally, combining these two schemes, we provide a computationally efficient and \textit{almost} optimal scheme to solve the channel allocation problem, and various results and insights on the variation of the optimal system performance with respect to different system parameters.
\end{abstract}

\begin{IEEEkeywords}
Device-to-device communication, resource allocation, LTE-A, stochastic geometry, interference management
\end{IEEEkeywords} 

\IEEEpeerreviewmaketitle

\section{Introduction}
\label{sec:1}
The unprecedented increase in the number of mobile users and their data demands has compelled the industry and academia to provide new approaches to further increase the network capacity. Media rich applications such as IPTV, mobile TV, high definition video streaming, and video conferencing, which currently account for more than 40\% of the internet traffic, are some of the disruptive innovations that can be supported by multicasting \cite{andrews2014will}. Compared to communicating with each receiver separately, multicast transmission has less signaling overhead, enhances spectral efficiency, and reduces transmission power consumption at the base station. Multicast in cellular networks is classified as: single-rate and multi-rate.  In single-rate multicast, the transmitter transmits to all the receivers at a common rate \cite{afolabi2013multicast}. For example, in \cite{liu2008dynamic}, authors formulate the sum throughput maximization problem in OFDMA-based multicast networks, and allocate the power and channel by adapting to receiver with the weakest link. While in multi-rate multicast communication \cite{tu2016efficient}, different receivers in a cluster may receive at different rate depending on their channel qualities. Although  more efficient, implementation and analysis of multi-rate multicast is much more complex than single-rate multicast.

In Long Term Evolution (LTE) networks, also known as single frequency networks, enhanced Multimedia Broadcast/Multicast Service (eMBMS) offers new opportunities to support value-added multicast applications such as weather forecasting, ticker-tape feeds, and local advertisements, which require the same chunk of data distributed to geographically proximate users \cite{lecompte2012evolved}. Moreover, eMBMS is efficient in quickly pushing the same content to many devices at the same time without interacting with the user. For example, popular content like podcast, advertisements and software upgrades can be cached and pushed to all the end-users during off-peak hours \cite{monserrat2012joint}. However, in single frequency networks, tight synchronization among base stations is needed. In addition, it is not an energy efficient solution as the base stations may need to transmit at the maximum power.

Underlay device-to-device (D2D) communication, which enables mobile devices in close proximity to bypass the base station (BS) and directly communicate with each other by reusing the spectrum, is being proposed as a new approach for cellular multicast \cite{tehrani2014device}. Being a short-range transmission technology, it may offer higher data rates, energy and spectrally efficient transmissions, and improved coverage \cite{lien20163gpp}. D2D multicast appears as a viable solution in such scenarios since a mobile device needs energy only when transmitting to other mobile devices in its group. Furthermore, in D2D multicast mobile devices communicate only with geographically proximate devices, thus yielding higher multicast bit-rate per unit of transmission energy compared to schemes where the BS multicasts to devices spread over large area. However, extensive deployment of underlay D2D multicast in a network may cause severe co-channel interference due to spectrum reuse, and rapid battery depletion of the multicasting D2D transmitter nodes combating the co-channel interference. Therefore, D2D multicast schemes are desired that efficiently manage the co-channel interference to maximize the network throughput.

Recently, D2D multicast in cellular networks has been considered in \cite{zhao2016resource,feng2018resource,meshgi2017optimal,yu2016d2d }. In \cite{zhao2016resource}, the problem of maximizing the number of accessed D2D groups while minimizing the total device transmission power for D2D underlay multicast communication is cast as mixed integer nonlinear programming (MINLP) problem and a joint power and channel allocation scheme is proposed to solve it. However, it only addresses the scenarios where the number of multicast groups (MGs) are less than the number of cellular users (CUs).  In \cite{feng2018resource}, authors consider social content sharing in cellular networks using D2D multicasting and propose a bipartite matching based channel allocation scheme to solve it. In \cite{meshgi2017optimal}, the problem of power and channel allocation for D2D multicast is formulated as a problem to maximize the sum throughput of CUs and MGs, which is further cast as an MINLP problem. However, the number of MGs per channel and the number of receivers per MG, are assumed to be fixed without any justification. In \cite{yu2016d2d}, the problem of overall energy minimization is formulated, and a flexible D2D multicast communication framework is provided to solve it.

In our previous work, we discuss some approaches for the sum rate maximization \cite{bhardwaj2015resource, bhardwaj2016resource, bhardwaj2017interference, bhardwaj2018channel} and energy-spectral efficiency tradeoff \cite{bhardwaj2018energy} in D2D multicast in underlay cellular networks.

Previously, stochastic geometry based approaches have been used for various D2D communication scenarios in underlay cellular networks \cite{ali2016modeling,guo2017device,lin2014modeling,lin2014spectrum,
chun2017stochastic,stefanatos2015operational,george2015analytical, yazdanshenasan2016poisson}. In \cite{ali2016modeling}, the locations of users are modeled as Poisson Point Process (PPP), and it is shown that the maximum spatial frequency reuse can be obtained if only a small percentage of possible D2D nodes are enabled. Authors in \cite{guo2017device} provide expressions for outage probability at the BS and a typical D2D user in a finite cellular network region, and demonstrate the impact of path-loss exponent on spectral efficiency. In \cite{lin2014modeling}, the authors provide the design, implementation, and optimization of overlay D2D-multicast, and compute various parameters, such as coverage probability of all D2D receivers, the optimal number of retransmissions for successfully delivering data packets. The work in \cite{lin2014spectrum} provides a tractable model to analyze the spectral efficiency behavior for underlay and overlay D2D communication in cellular networks and in \cite{chun2017stochastic}, an approach to evaluate the rate and bit error probability of D2D networks over generalized fading channels is provided. In \cite{stefanatos2015operational}, the authors provide operational conditions in which D2D communications helps in increasing the system performance, and show the dependence of these conditions on various system parameters, such as user density. In \cite{george2015analytical}, the D2D node distribution is modeled by a Poisson Dipole Process (PDP), wherein transmitters are distributed as a homogeneous PPP, with the corresponding receivers being located at a fixed distance from the transmitter. To reduce the impact of interference on D2D receivers by the CUs, \cite{yazdanshenasan2016poisson} proposes to have an exclusion zone around the cellular users, thus saving D2D links from excessive CUs interference. It is shown that the interference can be significantly reduced if the locations of D2D transmitters are modeled as a Poisson Hole Process (PHP).

In this work, we propose a stochastic geometry based approach to maximize the sum throughput of CUs and MGs for D2D multicast enabled underlay cellular networks, where the cellular users and D2D receivers are distributed according to a PPP and PHP, respectively, without any restriction on the numbers of CUs, MGs, and receivers in each MG. The main contributions of this work are as follows:
\begin{enumerate}
\item We introduce the hybrid network model consisting of both the CU and D2D nodes. Specifically, we use the homogeneous PPP to model the spatial distribution of CUs and D2D MGs. A circular exclusion zone is considered around each CU, and only those receivers of the multicast groups which exist outside the exclusion zones are included.
\item We allow multiple MGs to share a channel with the corresponding cellular user. Further, subject to the target SIR constraint, any number of receivers can exist in any MG with no restriction on D2D transmitter-receiver distance. 
\item We propose a joint channel and power allocation problem to find the optimal subset of MGs for each cellular channel and the optimal transmit powers of MGs allotted to a particular channel. We decompose the problem of allocating channels to MGs into two subproblems: selection of subsets of MGs to be assigned to the available channels (subset selection problem) and assignment of channels to these selected subsets (channel assignment problem.) Using this decomposition, it is established that the channel allocation problem has computational complexity exponential in both, the number of available cellular channels and the number of multicast groups.
\item For the subset selection problem, first we numerically establish that though subsets of MGs of any size may share a cellular channel, however, allowing only a small and \textit{almost} equal sized subsets for each channels closely approximates the optimal performance. In fact, this performance remains almost unchanged if we consider only equal sized subsets. Restricting the subset sizes to be \textit{almost} equal or equal significantly reduces the combinatorially large search space for the optimal solution with little degradation in the optimal performance. Finally, restricting the subset sizes to a fixed and equal number, allows us to construct a computationally efficient and practical scheme for the subset selection problem without overly compromising the optimal performance.
\item For the channel assignment problem, based on a novel idea to characterize the maximum interference among the subset of MGs to be assigned to a channel, we propose a computationally efficient \textit{almost} optimal scheme to allocate subsets of MGs to available channels.
\item Lastly, combining the two proposed computationally efficient schemes for subset selection and channel assignment problems, we construct a combined computationally efficient scheme for the channel allocation problem. We compare the performance of this combined scheme with the optimal performance with respect to the variation of different system parameters and conclude with some insights on the optimal design and operation of such cellular multicast system.
\end{enumerate}

\noindent\textit{Organization:} The manuscript is organized as follows. The system model is described in Section \ref{sec:2}. The detailed problem formulation is proposed and analyzed in Section \ref{sec:3}. The design and analysis of a computationally efficient and \textit{almost} optimal scheme for the channel allocation to MGs is described in Section~\ref{sec:gcas}. The numerical results are presented in Section \ref{sec:numRes}. Finally, Section \ref{sec:concl} concludes the manuscript and discusses ways to extend this work.

\section{System Model}
\label{sec:2}
We consider a D2D-enabled underlay cellular network that is modeled 
as a Poisson Hole Process (PHP) in a region $\mathcal{R}^2$ \cite{yazdanshenasan2016poisson}. Let $\mathcal{C} = \{1,2,\ldots,C\}$ denote the number of orthogonal uplink channels that are shared by D2D multicast groups (MGs) and cellular users (CUs). Sharing of uplink channels is considered as the downlink channels have more traffic load \cite{onireti2012energy}, and also as it is the BS that faces interference in uplink sharing mode, it may handle interference more effectively than a mobile device in the downlink. The spatial distribution of CUs and D2D MGs on the $k^{\textrm{th}}$ channel is modeled as homogeneous PPP, $\Pi_{c,k}$ with density $\lambda_{c}^k$, and $\Pi_{g,k}$ with density $\lambda_{g}^k$, respectively. In a homogeneous distribution, the MG receivers (MGRXs) may become very close to CU transmitters (CUTXs), and may suffer co-channel interference from a CU. Therefore, to protect the D2D MG receivers from co-channel interference created by the CUs, a circularly shaped exclusion zone of radius $D$ is considered around each CU. The region covered by the exclusion zones can be expressed as 
\begin{equation}
\mathcal{B}_D = \underset{ \textbf{x}_c \in \Pi_{c,k}}{\bigcup} \textbf{b}\left(\textbf{x}_c ,D\right), ~\textbf{b}(\textbf{x}_c,D) = \lbrace \textbf{x} \in \mathbb{R}^2: \vert \vert \textbf{x} - \textbf{x}_c \vert \vert <D\rbrace, \nonumber
\end{equation}
where $\textbf{b}(\textbf{x}_c ,D)$ denotes the circle of radius $D$ centered at the location $\textbf{x}_c$ of a cellular user. We assume that all D2D receivers which lie within the exclusion zone of any CU are not part of any MG\footnote{This is a pessimistic assumption. The system performance may improve if a D2D receiver can join an MG even if it is within the exclusion zone of some CU provided that MG does not share the channel with the said CU. However, we work with this assumption because it simplifies analysis without altering essential characteristics of the optimal solution, and our proposed schemes.}. Conversely, any receiver that is outside the exclusion zones of all CUs can join any MG, thus any MG may have any number of receivers. 
For formation of MGs in eMBMS,  initially it is assumed that all MG transmitters (MGTXs) transmit at the maximum power. Any receiver is connected to only that MGTX for which it has the maximum SNR, and it should be above certain threshold. We follow the same model for forming MGs.

Let ${\mathcal G}$ denote the set of MGs in the cell and $G = |{\mathcal G}|$. Similarly, let ${\mathcal C}$ denote the set of cellular channels in the cell and $C = |{\mathcal C}|$. Assume that each cellular channel is assigned to one and only one CU. Each channel is assumed to be Rayleigh faded and to follow a power law path-loss with exponent $\alpha$.  Therefore, the received power at the $j^{\textrm{th}}$ receiver from the $i^{\textrm{th}}$ serving node is $p_j = p_i h_{i,j}d_{i,j}^{-\alpha}$, whereas $d_{i,j}$ denotes the Euclidean distance between the $i^{\textrm{th}}$ and the $j^{\textrm{th}}$ nodes.  Let $u_g$ be a mobile user associated with the $g^\textrm{th}$ MG, and the corresponding set is denoted as $\mathcal{U}_g$, where cardinality of $\mathcal{U}_g$ determines the number of receivers in the $g^\textrm{th}$ MG ($|\mathcal{U}_g|=1$ corresponds to unicast communication.)

The transmit powers of a CU and a MG on the $k^{\textrm{th}}$ channel are denoted by $p_{c,k}$ and $p_{g,k}$, respectively, and the corresponding maximum values are $P_c$ and $P_G$, respectively. As we are considering the scenarios where a channel is shared by a CU and multiple MGs, the $r^{\textrm{th}} (r \in \mathcal{U}_g)$ D2D receiver experiences interference from the co-channel MGTXs and the CU. As the system model is interference limited, so instead of signal-to-interference and noise ratio (SINR) we consider the signal-to-interference ratio (SIR). The SIR at the $r^{\textrm{th}}$ receiver in the $g^{\textrm{th}}$ MG sharing the $k^{\textrm{th}}$ channel is denoted as:
\begin{equation*}
\label{eq:2}
\gamma_{g,r}^k =  \frac{p_{g,k} h_{g,r,k} d_{g,r}^{-\alpha}}
{p_{c,k} h_{c,r,k} d_{c,r}^{-\alpha} + \underset{g' \in \Pi_{g,k}}{\sum}p_{g',k} h_{g',r,k} d_{g',r}^{-\alpha}},
\end{equation*}
Let $I_{c,r,k} = p_{c,k} h_{c,r,k} d_{c,r}^{-\alpha} $ and $I_{g',g,k} = \sum_{g' \in \Pi_{g,k}} p_{g',k}$ $ h_{g',r,k} d_{g',r}^{-\alpha}$. 
The maximum possible transmission rate in MGs is determined by the channel conditions of the worst receiver \cite{meshgi2017optimal}. Therefore, the corresponding SIR is
\begin{equation*}
\gamma_g^k  = \min_{r \in \mathcal{U}_g} \left( \frac{p_{g,k}h_{g,r,k} d_{g,r}^{-\alpha}}{I_{c,r,k}  + I_{g',g,k}}\right)
\end{equation*}
An outage event for the $g^{\textrm{th}}$ MG occurs if the received SIR for any of its receiver is less than the minimum acceptable SIR, $\gamma^\textrm{th}_g$. The corresponding outage probability is given by the following lemma.
\begin{pavikl}
\label{lemma_1}
The outage probability of a receiver in a D2D MG distributed as PHP and communicating on the $k^\textrm{th}$ shared channel is
\begin{align*}
\textrm{Pr}(\gamma_g^k < \gamma_g^\textrm{th}) &= 1 - \mathcal{L}_{I_{c,r,k}} \left( \gamma_g^\textrm{th} p_{g,k}^{-1} d_{g,r}^\alpha \right) \times \mathcal{L}_{I_{g',g,k}} \left(\gamma_g^\textrm{th} p_{g,k}^{-1} d_{g,r}^\alpha\right) \\
&=  1- \mathcal{L}_1 \left( \lambda_c^k, p_{c,k}, D, \gamma_g^\textrm{th} p_{g,k}^{-1} d_{g,r}^\alpha \right) \times \mathcal{L}_0\left( \lambda_g^k, {p_{g,k}}, \gamma_g^\textrm{th} p_{g,k}^{-1} d_{g,r}^\alpha \right)
\end{align*}
where $\delta = 2/ \alpha,$ and with $\mathcal{Y}_1 =\gamma_g^\textrm{th} p_{g,k}^{-1} d_{g,r}^\alpha p_{c,k} D^{-\alpha} $
\begin{align*}
\mathcal{L}_0\left( \lambda_g^k, {p_{g,k}}, \gamma_g^\textrm{th} p_{g,k}^{-1} d_{g,r}^\alpha \right) &= \exp \bigg( - \lambda_g^k \frac{\pi^2}{2} p_{g,k}^{\frac{1}{2}} \sqrt{\gamma_g^\textrm{th} p_{g,k}^{-1} d_{g,r}^\alpha} \bigg), \\
\mathcal{L}_1 \left( \lambda_c^k, p_{c,k}, D, \gamma_g^\textrm{th} p_{g,k}^{-1} d_{g,r}^\alpha \right) &= \exp\Bigg(-\lambda_c^k \pi \Bigg\lbrace p_{c,k}^{\frac{1}{2}} \sqrt{\gamma_g^{\textrm{th}}} p_{g,k}^{-\frac{1}{2}} d_{g,r}^{\frac{\alpha}{2}} \bigg(\arctan\left(\sqrt{\mathcal{Y}_1}\right) + \frac{\sqrt{\mathcal{Y}_1}}{1+ \mathcal{Y}_1}\bigg)\Bigg\rbrace -  \frac{\mathcal{Y}_1 D^2}{ 1 + \mathcal{Y}_1}\Bigg)
\end{align*}
\end{pavikl}
\begin{IEEEproof}
The proof can be obtained by following the argument as in \cite[eq. 3.21,3.46]{haenggi2009interference}. 
\end{IEEEproof}
Similarly, the $k^\textrm{th}$ CU's transmission suffers interference, $I_{gc}$, from D2D MGs which are operating on the $k^\textrm{th}$ channel. Therefore, the SIR of a CU at the BS is given as
\begin{equation*}
\gamma_c^k = \frac{p_{c,k} h_{c,b} d_{c,b}^{-\alpha}}{ \underset{g \in \Pi_{g,k}}{\sum}p_{g,k} h_{g,b,k} d_{g,b}^{-\alpha}}
\end{equation*}
An outage event for the $k^{\textrm{th}}$ CU occurs if the received SIR for it at the BS is less than the minimum acceptable SIR, $\gamma^\textrm{th}_c$. The corresponding outage probability is given by the following lemma.
\begin{pavikl}
\label{lemma_2}
The outage probability of a CU transmitting on the $k^\textrm{th}$ shared channel is
\begin{align*}
\label{eq:5}
\textrm{Pr}(\gamma_c^k < \gamma_c^{\textrm{th}}) &= 1- \mathcal{L}_{I_{gc}} \left(\gamma_c^\textrm{th} p_{c,k}^{-1} d_{c,b}^\alpha\right) \nonumber \\
&= 1 - \mathcal{L}_0 \left( \lambda_g^k, p_{g,k}, \gamma_c^\textrm{th} p_{c,k}^{-1} d_{c,b}^\alpha\right),
\end{align*}
where
\begin{align}
\mathcal{L}_0 \left( \lambda_g^k, p_{g,k}, \gamma_c^\textrm{th} p_{c,k}^{-1} d_{c,b}^\alpha\right)= \exp \bigg( - \lambda_c^k \frac{\pi^2}{2} p_{g,k}^{\frac{1}{2}} \sqrt{\gamma_c^\textrm{th} p_{c,k}^{-1} d_{c,b}^\alpha} \bigg) \nonumber
\end{align}
\end{pavikl}
\begin{IEEEproof}
The proof can be obtained by following the same argument as in \cite[eq. 3.21,3.46]{haenggi2009interference}. 
\end{IEEEproof}

The basic notation used in this paper is summarized in Table \rm{I}.
\renewcommand{\arraystretch}{1}
\begin{table}[!t]
\caption{Major Notation} 
\label{Table_1}
{%
\begin{tabular}[width=\linewidth]{p{0.75in} p{6.0in}}
\hline
$\mathcal{C}$ & Set of cellular users, $N_c = \vert \mathcal{C}\vert $\\
$p_{c,k}$  & The transmission power of the $k^\textrm{th}$ CU\\
$P_c^{\max}$ & The maximum power that can be transmitted by any CU\\
$p_{g,k}$  & The transmission power of the $k^\textrm{th}$ MG transmitter\\
$P_g^{\max}$ & The maximum power that can be transmitted within any D2D MG\\
$D$ & Exclusion zone radius\\
$\gamma^k_c$ & The SINR received by eNB on the $k^{\textrm{th}}$ channel\\
$R_{k,c}^{\min}$ & The minimum data rate required by the $k^{\textrm{th}}$ CU\\  
$\mathcal{G}$ & Set of multicast transmitters, G = $\vert \mathcal{G}\vert $ \\ 
$\mathcal{G}_k$ & Set of MGs communicating on channel k\\
$\Pi_{c,k} $ & The spatial distribution of CUs on the $k^{\textrm{th}}$ channel with density $\lambda_c^k$\\
$\Pi_{g,k} $ & The spatial distribution of D2D MGs on the $k^{\textrm{th}}$ channel with density $\lambda_g^k$\\ 
$\mathcal{U}_g$ & Set of mobile receivers associated with the $g^\textrm{th}$ MG\\
$\Theta_c, \Theta_g$ & Outage thresholds for CUs and D2D receivers, respectively\\
\hline
\end{tabular}
}
\end{table}

\section{Problem Formulation}
\label{sec:3}
In order to ensure acceptable SIR to both CUs and D2D MG transmissions, thresholds on outage probabilities are included as follows:. 
\begin{equation}
\label{eq:7}
\textrm{Pr~}(\gamma_c^k < \gamma^{\textrm{th}}_c)  \leq \Theta_c, \mbox{ and } ~  \textrm{Pr~} (\gamma_g^k < \gamma_g^{\textrm{th}}) \leq \Theta_g,
\end{equation}
where $\Theta_c$ and $\Theta_g$ are the parameters that represent the outage thresholds for CUs and D2D receivers, respectively. The transmit power of the $g^\textrm{th}$ MG on the $k^\textrm{th}$ channel must be less than its possible maximum value, $P_G$. Thus, we have, $0 \leq p_{g,k} \leq P_G$. As the following lemma shows, we can further narrow down the range of possible values of $p_{g,k}$.
\begin{pavikl}
\label{lemma_feasibleRegion}
The feasible region for $p_{g,k}$ to satisfy outage constraints in \eqref{eq:7} is
\begin{equation*}
p_{g,k}^\textrm{low} \le p_{g,k} \leq p_{g,k}^{\textrm{high}},
\end{equation*}
where
\begin{align*}
p_{g,k}^\textrm{low} &=  \frac{2 \lambda_c^k \pi p_{c,k}\gamma_g^{\textrm{th}}  d_g^{\alpha}}{- \ln(1- \Theta_g) - \lambda_g^k \frac{\pi^2}{2}\sqrt{\gamma_g^{\textrm{th}}} d_g^{\frac{\alpha}{2}} + \gamma_g^{\textrm{th}} d_g^{\alpha} D^{2 - \alpha} \lambda_c^k \pi}\\
p_{g,k}^{\textrm{high}} &=  p_{c,k} \bigg(\frac{2 \ln (1 -  \Theta_c)}{\lambda_g^k \pi^2 \gamma_c^\textrm{th} d_{c,b}^{\frac{\alpha}{2}}} \bigg)^2
\end{align*}
\end{pavikl}
\begin{IEEEproof}
Please refer to Appendix \ref{appndx:feasible_region}.
\end{IEEEproof}
With $p_{g,k}^\textrm{low}$  and $p_{g,k}^{\textrm{high}}$ defined as above, we further define $p_{g,k}^{\textrm{inf}} =  \max\{0, p_{g,k}^\textrm{low}\}$ and $p_{g,k}^{\sup}= \min\{P_G, p_{g,k}^\textrm{high} \}$. Thus, the problem to maximize the average sum throughput per unit area of a D2D multicast enabled cellular network while satisfying outage thresholds, can be posed as the following optimization problem:
\begin{align*}
\mathbf{P1:}~~~ &\underset{a_{gk}, p_{g,k},p_{c,k}} {\max~}  \sum\nolimits_{k=1}^{C} \left(a_{gk} R_g^k + R_c^k \right)\\
\textrm{s.t. \- } &\text{C}_1: p_{g,k}^{\inf} \leq p_{g,k} \leq p_{g,k}^{\sup} \\
&\text{C}_2: 0 \leq p_{c,k} \leq  P_c \\
&\text{C}_3: a_{gk} \in \lbrace0,1 \rbrace, ~\forall g \in \mathcal{G}, k \in \mathcal{K} \\
&\text{C}_4: \sum\nolimits_{k \in \mathcal{K}} a_{gk} = q, ~\forall g \in \mathcal{G},
\end{align*}
where $R_g^k = \lambda_g^k B_w  \log_2 \left( 1+  \gamma_g^k \right) \text{Pr}\left(\gamma_g^k  \geq \gamma_g^\textrm{th} \right)$ and $R_c^k = \lambda_c^k B_w \log_2 \left(1 + \gamma_c^k \right) \text{Pr}(\gamma_c^k \geq \gamma_c^{\textrm{th}})$. Constrains $\text{C}_1$ and $\text{C}_2$ provide the feasible power regions for MGs and CUs, respectively. Constraint $\text{C}_3$ ensures that only one channel is allocated to a MG. Constraint $\text{C}_4$ limits the number of MGs per channel to $q$.

Problem P1 is essentially about allocating $G$ or fewer MGs to $C$ channels to maximize the sum throughput of CUs and MGs, subject to the transmit power constraints of CUs and MGs. We visualize this problem as consisting of two problems: (1) a combinatorial problem of channel allocation to MGs (selection and assignment of different subsets of $G$ MGs to different channels), and (2) a non-convex optimization problem of power allocation to CUs and the subsets of MGs assigned to the corresponding cellular channels to maximize the sum throughput. This leads to the following equivalent formulation of Problem P1.
\begin{align*}
\mathbf{P2:}~~~ &\max_{\substack{\{M_1, \ldots, M_C\}\\M_i \in 2^{\mathcal{G}}}} \max_{p_{g,k},p_{c,k}} \sum\limits_{k=1}^{C} \left(R_c^k+\sum\limits_{g \in M_i}R_g^k \right)\\
\textrm{s.t. \- } &\text{C}_1: p_{g,k}^{\inf} \leq p_{g,k} \leq p_{g,k}^{\sup} \\
&\text{C}_2: 0 \leq p_{c,k} \leq  P_c\\
&\text{C}_3: \cup_{i=1}^C M_i \subseteq \mathcal{G}\\
&\text{C}_4: M_i \cap M_j = \varnothing, \forall \, i, j \in \{1, \ldots, G\},
\end{align*}
where $2^{\mathcal{G}}$ is the power set of $\mathcal{G}$, the set of all MGs.

It should be noted that if $G \le C$, the optimal solution is obtained in terms of bipartite matching where a channel is assigned to the MG that achieves the highest rate on that channel and vice-versa. Also, if one or more channels are not assigned to any MG, then the sum-throughput can always be increased by moving MGs from the channels to which two or more MGs are assigned to these channels. Therefore, in the rest of the paper we discuss the case where $G > C$ and there are always $C$ non-empty subsets of MGs to be assigned to $C$ channels. Thus, the simplest assignment to consider is $(1,1,..., 1)$, where at least one MG is assigned to each channel. Therefore, Problem P2 leads to

\begin{align*}
\mathbf{P3:}~~~ &\max_{q \in \{C, \ldots, G\}}\max_{\substack{\{M_1, \ldots, M_C\} \in \Pi_q^C\\M_i \in 2^{\mathcal{G}} \setminus \varnothing}} \max_{p_{g,k},p_{c,k}} \sum\limits_{k=1}^{C} \left(R_c^k+\sum\limits_{g \in M_i}R_g^k \right)\\
\textrm{s.t. \- } &\text{C}_1: p_{g,k}^{\inf} \leq p_{g,k} \leq p_{g,k}^{\sup} \\
&\text{C}_2: 0 \leq p_{c,k} \leq  P_c\\
&\text{C}_3: \cup_{i=1}^C M_i \subseteq \mathcal{G}, |\cup_{i=1}^C M_i| = q\\
&\text{C}_4: M_i \cap M_j = \varnothing, \forall \, i, j \in \{1, \ldots, G\},
\end{align*}
where $\Pi_q^C$ denotes the set of all permutations of non-empty subsets of $q$ MGs over $C$ channels.

This problem formulation explicitly brings out the selection (where at least $C$ out of $G$ MGs are selected for assignment to $C$ channels) and assignment (where $C!$ permutations of $C$ subsets of selected MGs are considered) subproblems of the channel allocation problem, and the power allocation problem inherent in the joint channel and power allocation problem.

Though the power allocation problem is important and interesting in its own right, however for the sake of brevity and to keep the discussion focused on the subset selection and channel assignment subproblems, in the rest of this paper we do not address the power allocation problem. As for a given assignment of MGs to various cellular channels the power allocation problem, in general, is non-convex, therefore for the sake of this work it is sufficient to note that such problems can be addressed using various optimization techniques for such problems \cite{lee2015power,qian2009mapel}. Therefore, in the rest of the paper, we concern ourselves with exploring the nature of the optimal solutions of selection and assignment subproblems and to construct \textit{almost} optimal computationally efficient solutions of these subproblems.

\noindent\textit{Remark on notation:} The number of MGs selected to be assigned to $C$ channels is denoted as $[x_1, \ldots, x_C]$, with $\sum_{i=1}^C x_i \le G$. For example, $[3, 2, 2]$ denotes that one channel is shared by three MGs, and other two channels are shared by two MGs each. It should be noted that it does not mean that the first channel has three MGs assigned to it, and the second and third channels have two MGs assigned to them, respectively.

Before discussing the motivation and design of computationally efficient schemes to address the subset selection and channel assignment subproblems, the reader may consider an example in Appendix~\ref{appndx:subsec:ssp} that illustrates the computation of different combinations for the channel allocation problem.

\begin{pavikp}
The total number of combination of MG subsets to be considered for channel allocation problem are:
\begin{equation}
\label{eqn:no_of_cases}
\left[\sum_{\substack{g_1 = 1, \ldots, G-(C-1)\\ g_i = 1, \ldots, g_{i-1}, i \in \{2, \ldots, C\}}} \dfrac{{G \choose g_1}{{G-g_1} \choose g_2} \cdots {{G - \sum_{i=1}^{C-1} g_i} \choose g_C}}{\prod_{g=1}^{G-(C-1)} \#_g}\right]C!,
\end{equation}
where $\#_g$ denotes the number of channels with $g$ MGs.
\end{pavikp}
\begin{IEEEproof}
As per Problem P3, $q$ goes from $C$ to $G$. One way to consider all possible $C$ non-empty subsets for each value of $q$ is to let one channel be assigned $g_1$ MGs, where $g_1$ goes form $1$ to $G-(C-1)$ MGs. Then, let another channel be assigned $g_2$ MGs where $g_2$ goes from $1$ to $g_1$, and so on upto the last channel. However, some channels may be degenerate in that they may have the same number of MGs assigned to them. To compensate for over-counting the number of subsets in such cases, we need to divide the total number of combinations for each value of $\{g_1, g_2, \ldots, g_C\}$ by the product of number of channels with a given number of MGs. This explains the term in the square brackets in the proposition which provides the solution for the subset selection subproblem.

The $C!$ factor outside the brackets corresponds to $C!$ permutations that need to be considered for each combination of MG subsets for the channel assignment subproblem.

The multiplication of these two factors gives the total number of combinations to be considered for the channel allocation problem.
\end{IEEEproof}

In Appendix~\ref{appndx:complexity}, we provide a lower bound for number of possible combinations to be considered for the channel allocation problem in Equation~\eqref{eqn:no_of_cases} and prove that it is at least exponential in both $C$ and $G$.

In the next section, we address the construction of a computationally efficient scheme for the channel allocation problem by first constructing such schemes for MG selection and channel assignment subproblems, and then cascading them together.

\section{Computationally Efficient and ``Almost'' Optimal Multicast Group Channel Allocation Scheme}
\label{sec:gcas}
As discussed in the last section, the MG channel allocation problem can be considered as consisting of two subproblems of MG subset selection and channel assignment to these subsets. Therefore, in this work we propose to construct a computationally efficient and \textit{almost} optimal scheme for MG channel allocation problem by constructing such schemes for the MG subset selection and the channel assignment subproblems, and cascading those two schemes together. In the next two subsections, we discuss the design and analysis of such schemes for these two subproblems, respectively.

\subsection{Computationally Efficient and ``Almost'' Optimal Multicast Group Selection}
\label{subsec:gss}
As established in the last section any non-empty subset of any size of MGs can be assigned to any of $C$ channels, resulting in exponentially large number of such possible assignments. However, given analytical intractability of the problem, we undertook an extensive numerical exploration of the nature of the optimal solution of this subproblem. Though in principle any number of MGs can be assigned to any channel, however our numerical experiments in Section~\ref{sec:numRes} have established that \textit{almost} optimal sum throughput is achieved when \textit{almost} equal number of MGs (specifically, the number of MGs assigned to any two channels do not differ by more than one) are assigned to each channel. In other words, though the optimal solution of Problem P3 is obtained by selecting $C$ non-empty subsets of MGs over all possible subsets of all possible sizes, our numerical exploration suggests that \textit{almost} optimal performance is achieved when this search for subsets of MGs is carried out only over all subsets such that the sizes of no two subsets differ from each other by more than one. Such subsets which result in the performance closest to the optimal performance are called \textit{almost} equal subsets. Let us call the scheme to compute such subsets \textit{almost} Equal.

This observation allows us to substantially cut down the number of subset selections to consider and develop a more efficient computational scheme to achieve \textit{almost} optimal sum throughput while solving the following optimization problem:
\begin{align*}
\mathbf{P4:}~~~ &\max_{q \in \{C, \ldots, G\}}\max_{\substack{\{M_1, \ldots, M_C\} \in \Pi_q^C\\M_i \in 2^{\mathcal{G}}}} \max_{p_{g,k},p_{c,k}} \sum\limits_{k=1}^{C} \left(R_c^k+\sum\limits_{g \in M_i}R_g^k \right)\\
\textrm{s.t. \- } &\text{C}_1: p_{g,k}^{\inf} \leq p_{g,k} \leq p_{g,k}^{\sup} \\
&\text{C}_2: 0 \leq p_{c,k} \leq  P_c\\
&\text{C}_3: \cup_{i=1}^C M_i \subseteq \mathcal{G}\\
&\text{C}_4: M_i \cap M_j = \varnothing, \forall \, i, j \in \{1, \ldots, G\}\\
&\text{C}_5: \max |M_i| - \min |M_j| \le 1, i, j \in \{1, \ldots, C\}, i \neq j
\end{align*}

The example in Appendix~\ref{appndx:subsec:aessp} demonstrates how the complexity of the subset selection problem is reduced when we consider \textit{almost} equal number of MGs in each subset.

As the scenarios where only an equal number of MGs are assigned to each channel are a subset of the scenarios with \textit{almost} equal number of MGs for each channel, in Appendix~\ref{appndx:complexity}, we provide a lower bound for number of possible combinations to be considered for the channel allocation problem corresponding to Problem P4 and prove that it is at least exponential in both $C$ and $G$.

Our numerical simulations in Section~\ref{sec:numRes} further establish that even allocating equal number of MGs to each cellular channel allows us to construct more efficient schemes while still tightly approximating the optimal sum throughput. In other words, instead of searching over all subsets whose sizes do not differ from each other by more than one, \textit{almost} optimal performance may still be achieved when this search for subsets is carried out only over equally sized subsets. Such a scheme solves a modified version of Problem P4 with constraint $\text{C}_5$ therein replaced by the following constraint: $\text{C}_5: \max |M_i| - \min |M_j| = 0, i, j \in \{1, \ldots, C\}, i \neq j$, as given below. Further, as we establish in Appendix~\ref{appndx:complexity}, such a scheme also provides lower bounds to the computational complexities for the optimal schemes to solve Problems P3 and P4. Let us call the scheme to compute such subsets Equal.
\begin{align*}
\mathbf{P5:}~~~ &\max_{q \in \{C, \ldots, G\}}\max_{\substack{\{M_1, \ldots, M_C\} \in \Pi_q^C\\M_i \in 2^{\mathcal{G}}}} \max_{p_{g,k},p_{c,k}} \sum\limits_{k=1}^{C} \left(R_c^k+\sum\limits_{g \in M_i}R_g^k \right)\\
\textrm{s.t. \- } &\text{C}_1: p_{g,k}^{\inf} \leq p_{g,k} \leq p_{g,k}^{\sup} \\
&\text{C}_2: 0 \leq p_{c,k} \leq  P_c\\
&\text{C}_3: \cup_{i=1}^C M_i \subseteq \mathcal{G}\\
&\text{C}_4: M_i \cap M_j = \varnothing, \forall \, i, j \in \{1, \ldots, G\}\\
&\text{C}_5: \max |M_i| - \min |M_j| = 0, i, j \in \{1, \ldots, C\}, i \neq j
\end{align*}

The example in Appendix~\ref{appndx:subsec:essp} demonstrates how the complexity of the subset selection problem is reduced further when we consider equal number of MGs in each subset.

In Appendix~\ref{appndx:complexity}, we provide an expression for number of possible combinations to be considered for the channel allocation problem corresponding to Problem P5 and prove that it is exponential in both $C$ and $G$.

A computationally efficient scheme that achieves \textit{almost} optimal performance can be constructed if instead of searching over all possible equally sized subsets, equally sized subsets of only one particular fixed size are used. Let us call this scheme \textit{fixed} Equal. This scheme is also practically relevant for cellular networks that support underlay D2D multicast. In such networks, for a given numbers of cellular channels and MGs requesting channel allocation, the size of \textit{almost} optimal equally sized subsets can be precomputed and stored in a table. Then, during the network operation when a certain number of MGs request channel allocation, the size of \textit{almost} optimal equally sized subset is read-off this table, the corresponding number (polynomial in $G$) of subsets are constructed, and the resulting channel assignment subproblem, and power allocation problem are solved.

\subsection{Computationally Efficient and ``Almost'' Optimal Multicast Group Channel Assignment}
\label{subsec:cas}
The solution of the subset selection subproblem provides $C$ non-empty subsets of MGs which need to be assigned to $C$ channels to find the optimal channel assignment for these MGs. This requires the consideration of all $C!$ permutations of these subsets over $C$ channels as discussed earlier in Section~\ref{sec:3}. This makes the assignment problem computationally intractable with respect to the number of channels $C$. However in this section, assuming each MGTX in the selected subsets to transmit at the maximum transmission power, we propose a computationally efficient algorithm to assign channels to these subsets as follows:

Let us assume that all MGTXs in the selected subsets are transmitting at the corresponding maximum transmission power\footnote{This scheme only assumes that each MGTX in the selected subsets is transmitting at the corresponding maximum power for the sake of channel assignment, but no power allocation is actually made, which is subsequently made at the power allocation stage.}. The algorithm proceeds in three major stages as follows:

\begin{enumerate}
\item For each CU, we first check if its received SIR at the BS is above its corresponding decoding threshold for each MGTX. If there is no MGTX that can meet the decoding threshold, then the corresponding channel is declared to not available for sharing with MGs.
\item For all selected subsets of MGs, for each MG compute the sum of interference from all other MGTXs in the subset and each CU at each receiver of the MG. Find the maximum sum interference over all receivers in the MG and thus declare the worst receiver and the corresponding sum interference. After having done this for all MGs in the subset, find the maximum sum interference over all MGs in the subset for each CU.
\item Stage 2 essentially results in a two-dimensional matrix with rows labeled by CUs and columns by the selected subsets. For each subset, we assign the channel that minimizes the worst sum interference for that subset. In case a channel minimizes the worst sum interference for more than one subsets, then it is assigned to the subset for which it leads to the smallest value.
\end{enumerate}

Stage 1 has complexity proportional to $C$. Stage 2 requires computation of the worst sum interference for each of $C$ subsets for each of $C$ channels, thus has complexity proportional to $C^2$. Finally, Stage 3 requires sorting the worst sum interference for each of $C$ channels, which has the worst-case complexity of $C^2$ and as this is to be done for each of $C$ subsets, the overall complexity is proportional $C^3$. Thus, the complexity of the proposed scheme is proportional to $C^3$.

In Algorithm~\ref{algo:Almost_optimal}, we provide the detailed pseudo-code of MUSCA algorithm that implements this design of a computationally efficient and \textit{almost} optimal scheme for the MG subset channel assignment problem.

By fixing the number of MGs per channel to a predefined and equal number (as in \textit{fixed} equal scheme defined in the last subsection), we first construct a polynomial (in $G$) number of combinations of MG subsets and then assign channels to them using MUSCA algorithm. This provides a computationally efficient scheme for the channel allocation problem. We refer to such a scheme as ``fixed-MUSCA'' algorithm.

\begin{algorithm}[!t]
\caption{Multicast group Subset Channel Assignment (MUSCA) Algorithm}
\label{algo:Almost_optimal}
\DontPrintSemicolon
\KwIn{$K,C,G,p_g = P_G,p_c = P_C,q, \mathcal{A}_c = \emptyset, \mathcal{S}$, where $\mathcal{S}$ denotes the 
number of MGs sets, $\vert \mathcal{S}\vert = K$ \\
\textbf{Output:} $R$}
\Begin
{
\textbf{Stage 1 (Feasibility Test):} \\
\For{k = 1: K}
{
\For{g = 1: $\mathcal{G}$}
{
Find $\gamma_{c}^k (k,g)$  
}
\If{$ \gamma_{c}^k(k,g),\forall_{g \in \mathcal{G}} < \gamma_c^{\textrm{th}}$}
{
$\mathcal{V}_k =  \mathcal{V}_k \cup \{k\}$ /only CU transmits on these channels/
}
\Else
{
$\mathcal{A}_k =  \mathcal{A}_k \cup \{k\}$ /number of available channels for sharing/
}
}
\textbf{Stage 2 (Maximum Sum Interference):}\\

\For {i = 1 : $\mathcal{S}$} 
{
\For {k = 1 : $\mathcal{A}_k$} 
{
\For {g = 1 : $\mathcal{G}$} 
{
\For {r = 1 : $RXs$} 
{
$\mathcal{I}_{c,g} = \max_{\substack{\{j \in C\}}} \max_{\substack{\{ g \in \mathcal{G}\}\\ \{r \in G_{RXs}\}}} \left(\sum I_{g',g} + I_{c,r}\right)$
}
}
}
}
\textbf{Stage 3 (Channel Assignment):} \\
\For {i = 1 : $\mathcal{S}$} 
{
$\mathcal{L}_{S,c} =  \argmin_{C} \mathcal{I}_{c,g}$
}
}
\end{algorithm}

\section{Numerical Results}
\label{sec:numRes}
In this section, we first discuss numerical simulation results that motivated the development of \textit{almost} optimal yet computationally efficient algorithms to solve the subset selection subproblem. Then, we discuss the performance of MUSCA algorithm for efficiently solving the assignment subproblem. Finally, we present results comparing the performance of fixed-MUSCA algorithm for channel allocation with the optimal results against variation of various system parameters. 

\textit{General Settings:} A D2D enabled cellular network of radius $R$ is considered, in which  CUs are uniformly distributed\footnote{Subject to knowing the number of nodes, the PPP is equivalent to the uniform distribution \cite{chiu2013stochastic}}. An exclusion zone of radius $D$ is assumed around each CU, and only those D2D receivers are enabled which lie outside all such exclusion zones. The main simulation parameters are listed in Table \ref{table:parameter}. It should be noted that though we have mostly considered only the network scenarios with $C=3$ and $G=7$, similar behavior is observed for networks with different values of $C$ and $G$ as long as $G > C$, as illustrated by some results.

\renewcommand{\arraystretch}{0.9}
\begin{table}[!t]
\setlength\extrarowheight{2.5pt}
\caption{Simulation Parameters} 
\centering
\begin{tabular}[width=3.5in]{|l|c|}
\hline
No. of cellular channels (C) & 3\\
\hline
No. of multicast groups (G)  & 7\\
\hline
Path loss exponent `$\alpha$' & 4 \\
\hline
$P_c$, $P_G$ & 30 dBm \\
\hline
$\gamma_g^\textrm{th}$ & 25 dB \\
\hline
No. of network scenarios for averaging & 500 \\
\hline 
\end{tabular} 
\label{table:parameter}
\end{table}

\subsection{``Almost'' optimal and computationally efficient solution for the subset selection subproblem}
\label{subsec:selection}
\begin{figure}
\centering
\includegraphics[width=0.9\hsize]{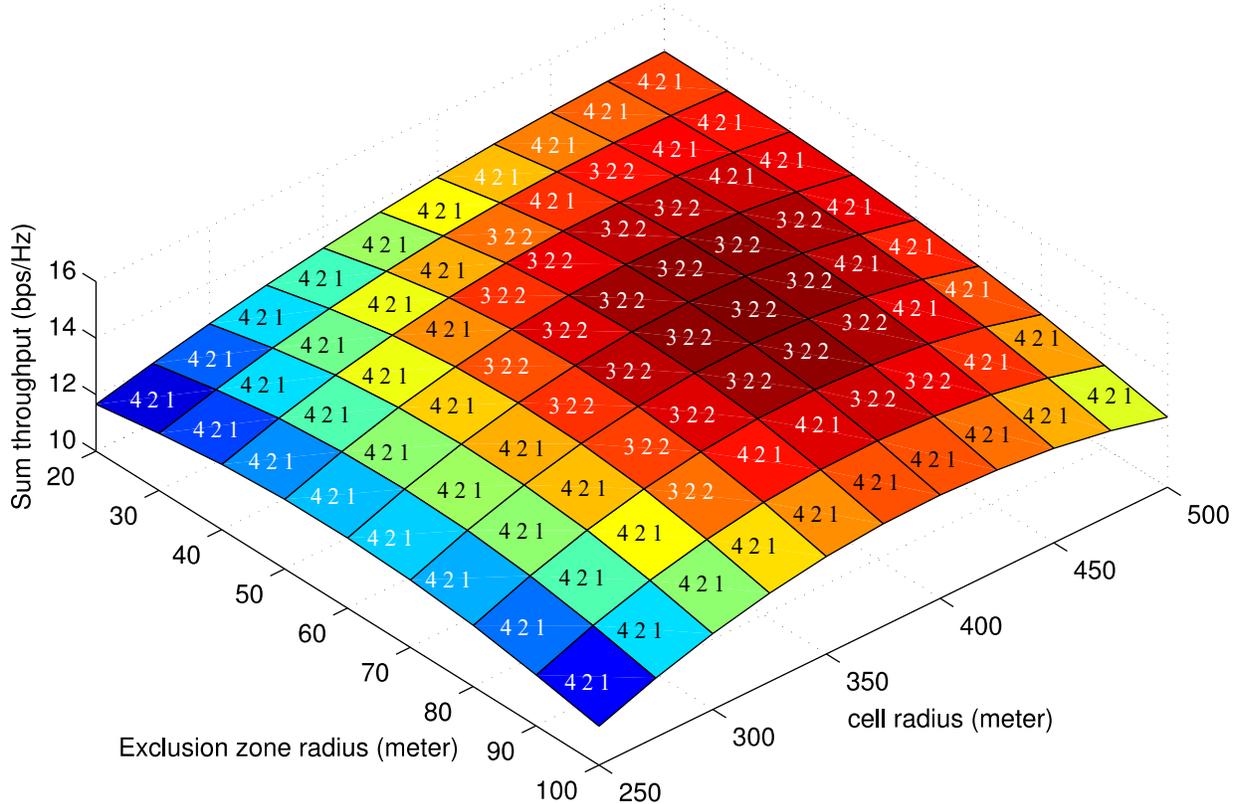}
\caption{The sum throughput as a function of exclusion zone radius and cell radius with $R_c^{\min} = 6$ bps/Hz and $\lambda_g = 2e^{-5}$.}
\label{fig:sum_thru}
\end{figure}

Figure~\ref{fig:sum_thru} depicts the sum throughput as a function of the exclusion zone radius, $D$, for different values of the cell radius, $R$. The value of $D$ is increased in the range of 20 to 100 meters, with a step size of 10 meters. The value of $R$ is increased in the range of 250 to 500 meters, with a step size of 50 meters. It can be observed that for each value of $R$, the sum throughput initially increases with increase in exclusion zone radius and then it decreases. This can be explained as follows: for smaller values of $D$, the D2D MGs receivers face severe interference from CU transmitters, therefore, they have low SIR, and consequently lower contribution to the sum throughput. For larger values of $D$, the number of D2D receivers which lies outside the exclusion zone are less, that is, $\vert\mathcal{U}_g\vert$ is less. This leads to decrease in D2D MGs' contribution to the sum throughput.

Also, the figure manifests the effect of cell radius on the sum throughput. This can be explained as follows: for a given value of $D$ and smaller values of $R$, the distance between the co-channel D2D MGs receivers is low, therefore, the interference among D2D MGs receivers is high, which leads to lower values of sum throughput. While for larger values of $R$, the receivers are far from their corresponding MGs transmitters, and to fulfill their SIR requirements, MG transmitters need to transmits at higher power, which may not be allowed as it may conflict with CUs' uplink outage constraints. Even in the cases where such high transmit powers of MG transmitters are allowed, those result in lower uplink SIR for CUs. Therefore, for every $D$ there exist an optimal value of $R$, for which  the interference between D2D MGs receivers is low, all receivers have a high value of SIR, and yet CUs' uplink rates are not overly compromised, which results in higher contribution to the sum throughput. Figure \ref{fig:sum_thru} also depicts for each ($R$, $D$) pair, the combination of MGs which contribute the maximum throughput for the highest number of times in 500 network scenarios, and it is observed that, [3, 2, 2] is the optimal combination of MGs for maximum value of the sum throughput. 

\begin{figure}[!t]
\centering
\includegraphics[width=0.7\hsize]{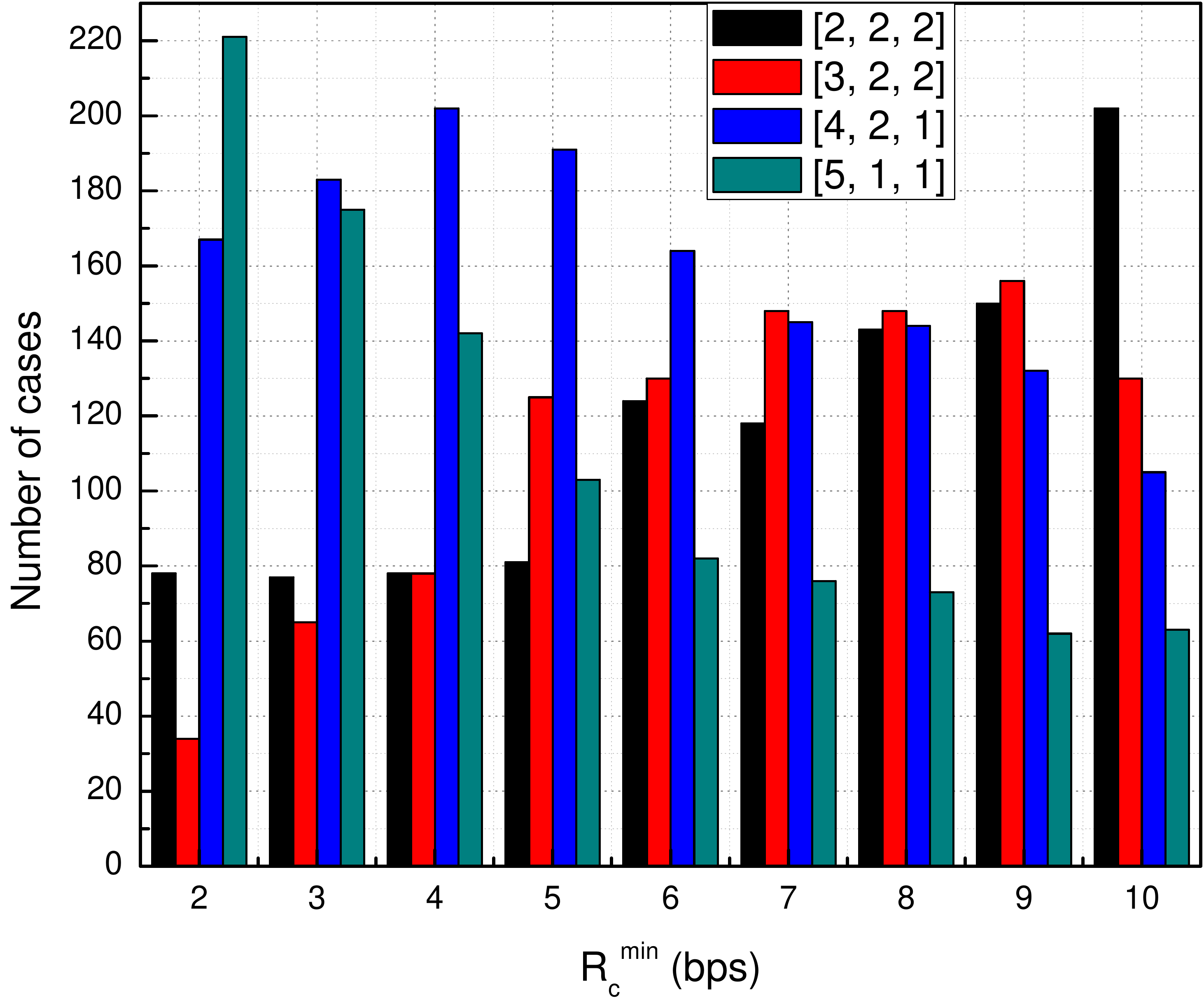}
\caption{Number of instances of different combinations out of 500 instances as a function of cellular users thresholds for $R$ = 500, $D$ = 50.}
\label{fig:no_of_cases_vs_rc_min}
\end{figure}

\begin{figure}[!h]
\centering
\includegraphics[width=0.7\hsize]{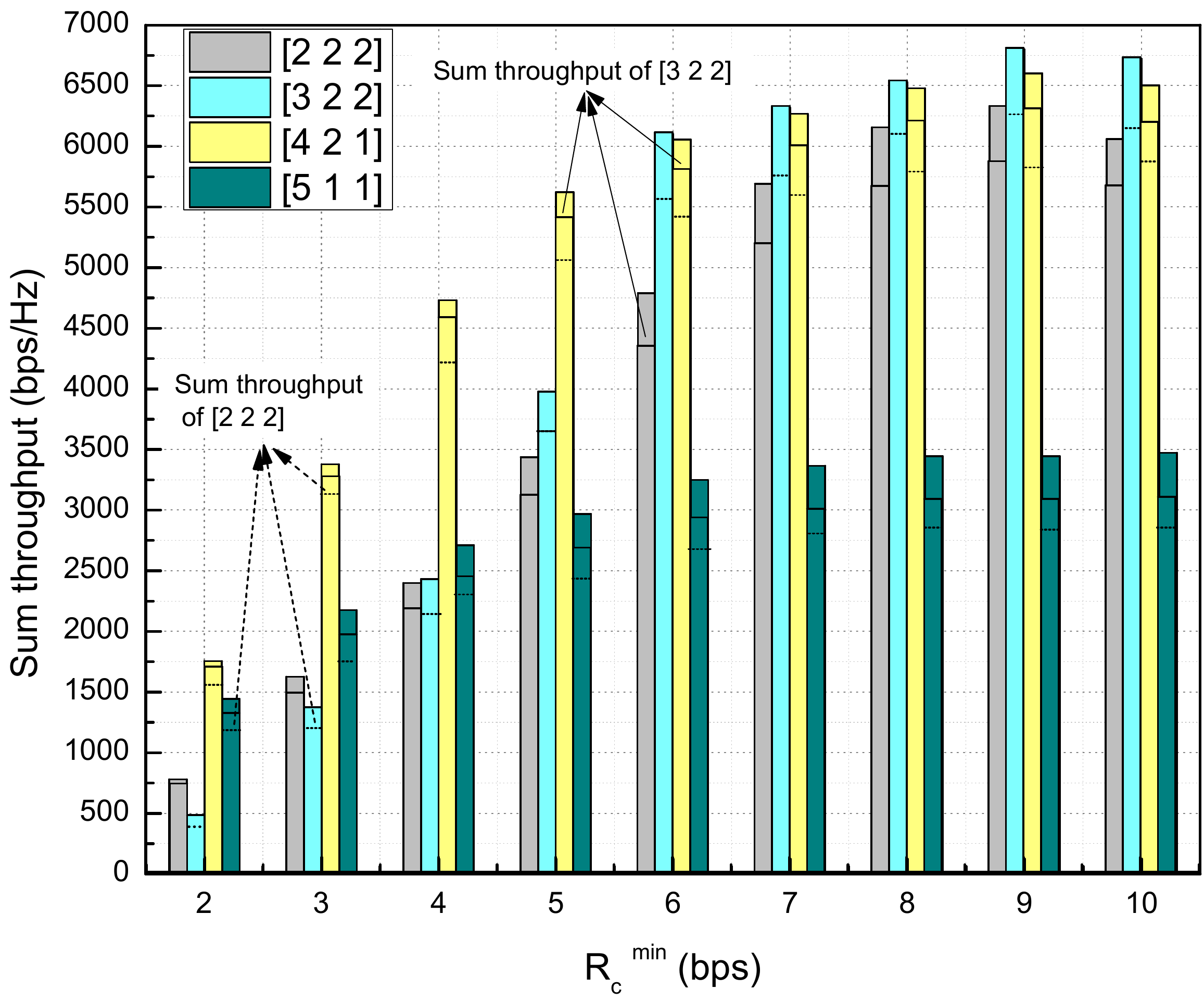}
\caption{The sum throughput of each combination as a function of cellular users thresholds for $R$ = 500, $D$ = 50.}
\label{fig:sum_thru_vs_rc_min}
\end{figure}

\begin{figure}[!t]
\centering
\includegraphics[width=0.7\hsize]{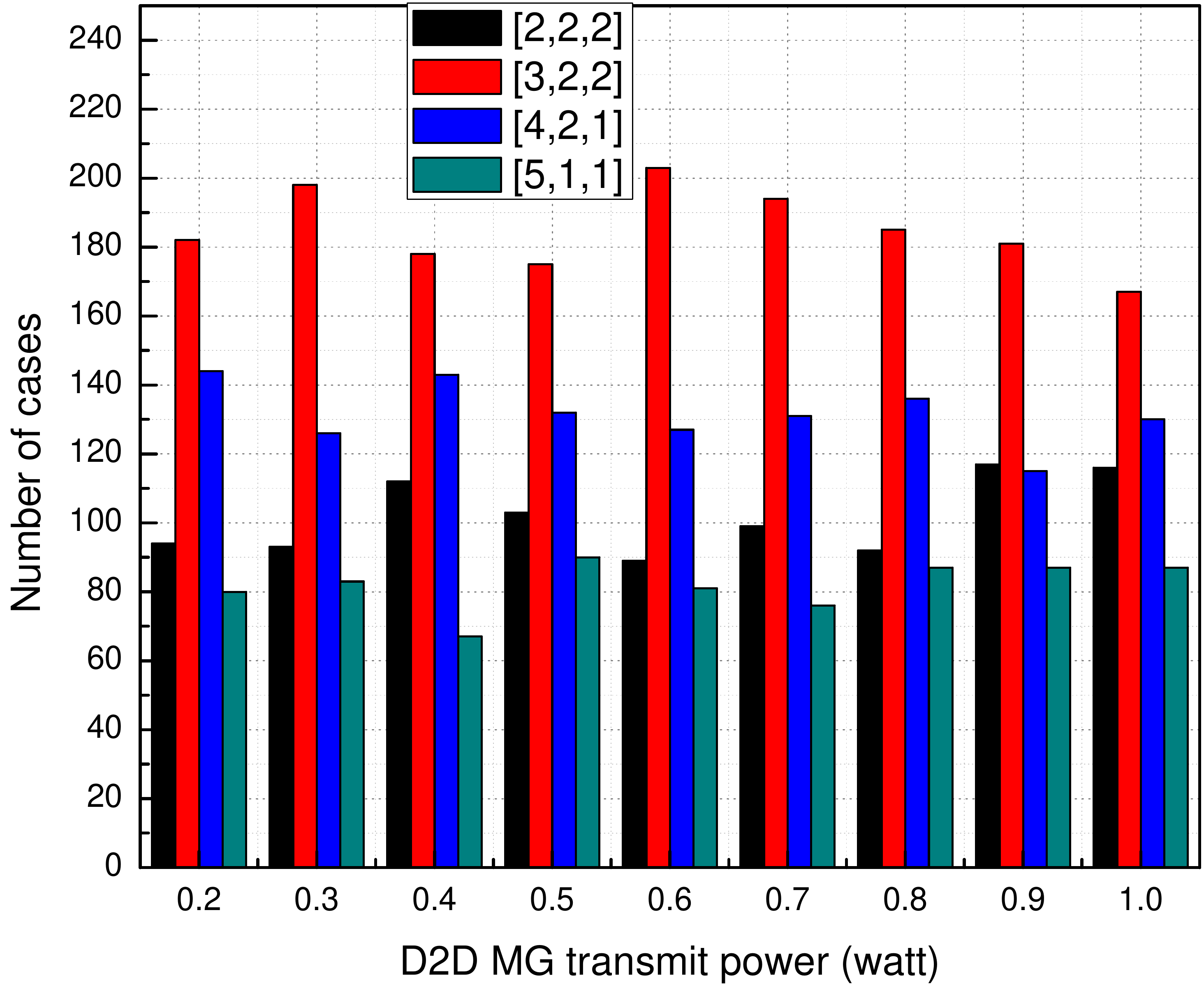}
\caption{Number of instances of different combinations out of 500 instances as a function of D2D MG transmit power for $R$ = 500, $D$ = 50.}
\label{fig:no_of_cases_vs_tx_pwr}
\end{figure}

\begin{figure}[!h]
\centering
\includegraphics[width=0.7\hsize]{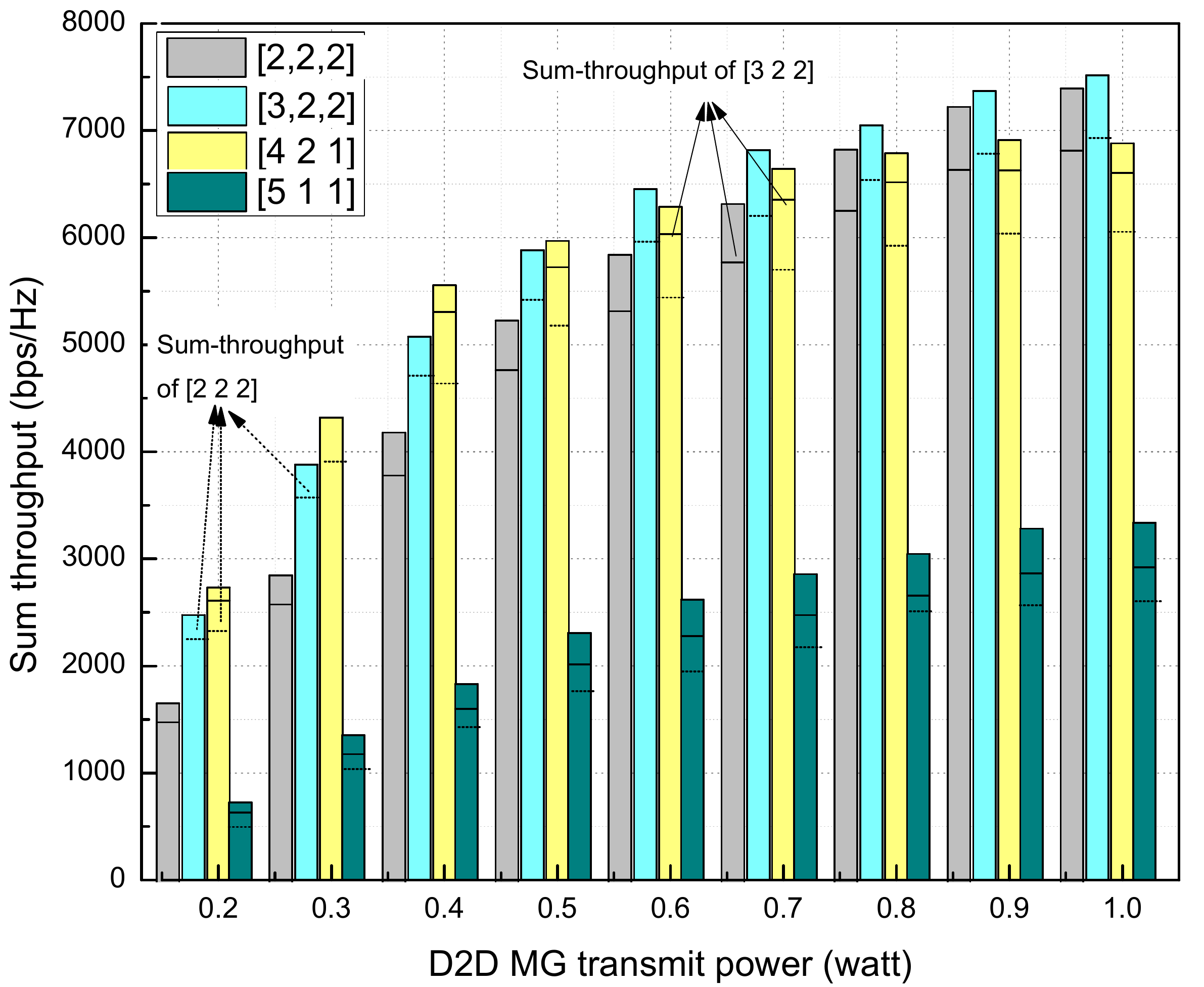}
\caption{The sum throughput of each combination as a function of D2D MG transmit power, $R$ = 500, $D$ = 50.}
\label{fig:sum_thru_vs_tx_pwr}
\end{figure}

In Figure~\ref{fig:no_of_cases_vs_rc_min}, for each value of $R_c^{\textrm{min}}$ we plot the number of instances (out of 500 network instances) for which a given MG combination is the optimal solution for Problem P3. Then in Figure~\ref{fig:sum_thru_vs_rc_min}, we plot the corresponding total sum throughput for each of these combinations for each value of $R_c^{\textrm{min}}$. Similarly, in Figure~\ref{fig:no_of_cases_vs_tx_pwr}, for each value of MG transmit power we plot the number of instances (out of 500 network instances) for which a given MG combination is the optimal solution for Problem P3. Then in Figure~\ref{fig:sum_thru_vs_tx_pwr}, we plot the corresponding total sum throughput for each of these combinations for each value of MG transmit power.

We can make multiple observations about these plots. First, though eleven MG combinations are possible solutions of Problem P3 in this case ($C=3, G=7$), only four combinations actually appear in all network scenarios for each value of $R_c^{\textrm{min}}$ and MG transmit power. Second, for all network instances for which some particular MG combination is optimal, using \textit{almost} equal combination [3, 2, 2] leads to no significant loss in the total sum throughput for those instances (the maximum loss is no more than 0.48 dB and 0.42 dB, respectively). The total throughput achieved by [3, 2, 2] in those instances where other combination is optimal is marked by a solid line in the corresponding bar in Figures~\ref{fig:sum_thru_vs_rc_min} and \ref{fig:sum_thru_vs_tx_pwr}. Third, for all network instances for which some particular MG combination is optimal, using equal combination [2, 2, 2] leads to no significant overall loss in the total sum throughput, though the loss is greater than it is for \textit{almost} equal combination [3, 2, 2] (the maximum loss is no more than 0.60 dB and 0.82 dB, respectively.) The total throughput achieved by [2, 2, 2] in those instances where other combination is optimal is marked by a dashed line in the corresponding bar in Figures~\ref{fig:sum_thru_vs_rc_min} and \ref{fig:sum_thru_vs_tx_pwr}.

\textit{Discussion:} In this problem scenario ($C=3, G=7$), for each network instance we have to consider eleven MG combinations ([1, 1, 1], [2, 1, 1], [2, 2, 1], [2, 2, 2], [3, 1, 1], [3, 2, 1], [3, 2, 2], [3, 3, 1], [4, 1, 1], [4, 2, 1], [5, 1, 1]) to search for the optimal solution of the MG selection problem. However, searching for the optimal combination over almost equal sized combinations reduces the search to only five combinations ([1, 1, 1], [2, 1, 1], [2, 2, 1], [2, 2, 2], [3, 2, 2]). This search reduces to just two combinations ([1, 1, 1], [2, 2, 2]) when we restrict ourselves to only equal sized combinations. This combined with our observations above provides a justification for successive approximation of Problem P3 by Problems P4 and P5, respectively, because each successive problem provides a close approximation to the optimal result of the previous one, but with fewer MG combinations to consider and thus lower corresponding complexity with respect to $G$.

\begin{figure}[!t]
\centering
\includegraphics[width=0.7\hsize]{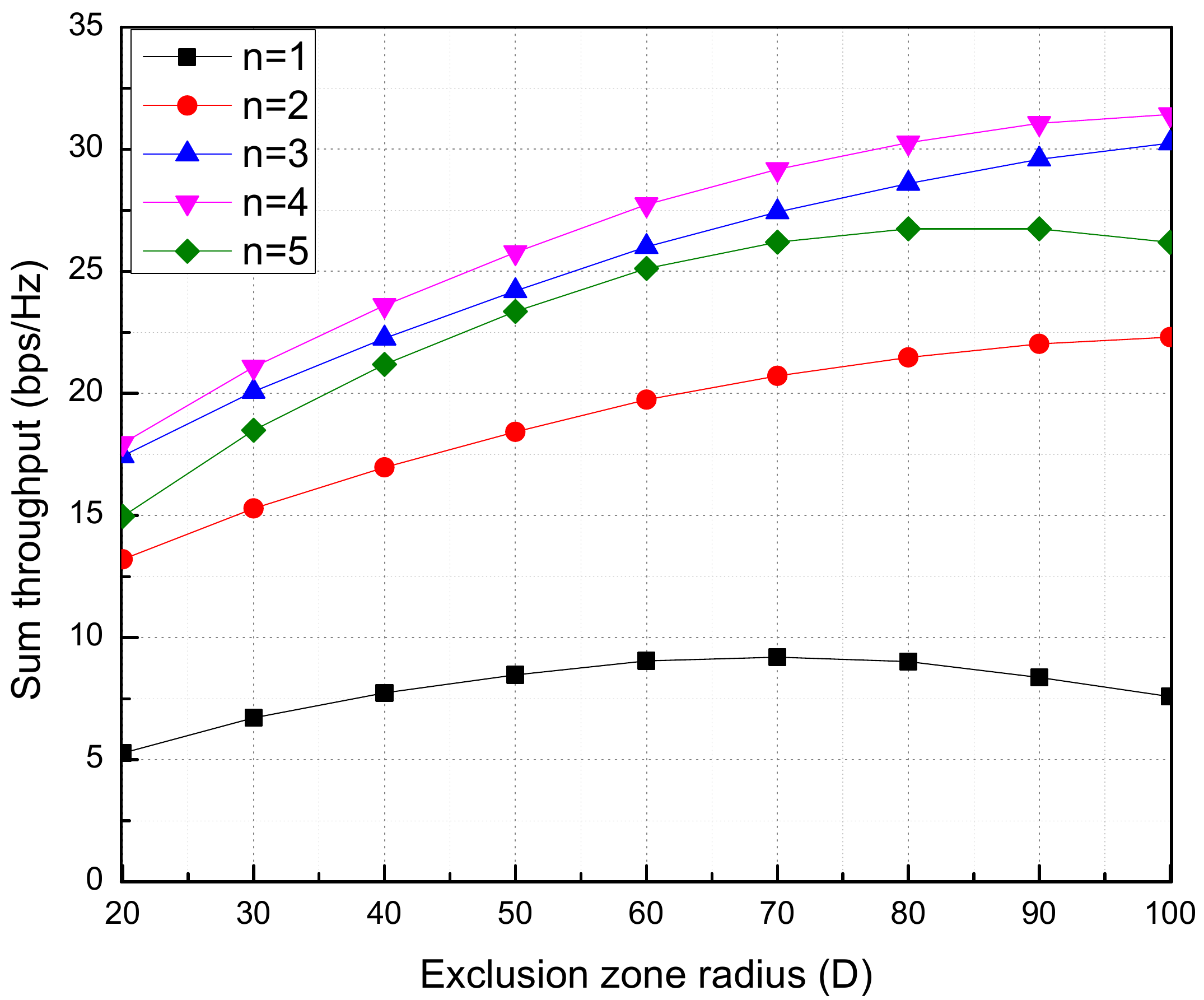}
\caption{The sum throughput as a function of exclusion zone radius with different number, $n$, of MGs per channel, $R$ = 500.}
\label{fig:n_var}
\end{figure}

In Figure~\ref{fig:n_var} the sum throughput is plotted for different number of MGs per channel against exclusion zone radius, $D$, and fixed cell radius, $R$, where the maximum sum throughput is attained for $n=4$ MGs per channel. It can be observed that the sum throughput first increases with increase in $n$, however after a certain number of MGs per channel, the sum throughput starts decreasing. This can be explained as follows: when we increase the number of MGs per channel, the contribution of each MG to the sum throughput increases along with mutual interference. Initially, increase in mutual interference is compensated by increasing MG transmit powers, however after a certain number of MGs per channel, the increase in mutual interference cannot be compensated by increasing power due to the constraint on maximum transmit power and the sum throughput starts to decrease with increasing number of MGs per channel beyond this.

\textit{Discussion :} Even when only equal sized MG combinations are allowed, the worst-case complexity of the selection subproblem remains exponential in $G$. However, increasing the number of MGs per channel does not monotonically increase the sum throughput: the sum throughput first increases with increasing number of MGs for each channel, reaches its peak for some particular number of MGs per channel (this number only depends on $C$ and $G$ when all other system parameter are fixed), and then it decreases with increasing number of MGs per channel. This motivates the argument at the end of Section~\ref{sec:gcas} for using only fixed and equal number of MGs per channel for given $C$ and $G$. When the number of MGs per channel is fixed, then the total number of combinations to consider is only a polynomial in $G$, leading to a computationally efficient solution for the selection subproblem without any significant loss with respect to the optimal sum throughput.

Table~\ref{table_comparison} illustrates a comparison of Optimal, \textit{almost} Equal, Equal and \textit{fixed} Equal subset selection schemes, for different combination of exclusion zone and cell radii for different maximum number of MGs per channel. In each table, we mark in boldface the value of $n$ for which the performance of \textit{fixed} Equal scheme is evaluated. It can be observed that for all schemes (except Equal), the sum throughput increases with increase in $n$ upto certain number, then it starts decreasing. The same behavior is observed for Equal as well, but except in one case ($D=100, R=250$), the corresponding peak occurs at $n=6$ or $7$ and we do not depict it here.

\begin{table}[!t]
\centering
\renewcommand\multirowsetup{\centering}
\caption{Comparison of optimal, Almost equal, Equal and Fixed equal schemes}
\footnotesize
\begin{tabular}{lSSSSSSSSSS}
\toprule
($D, R$) & \multicolumn{5}{c}{(20,250)} & \multicolumn{5}{c}{(20,500)}\\
\cmidrule(r){2-6}\cmidrule(l){7-11}
 & {n=1} & {n=2}  & {n=3} & $\mathbf{n=4}$ & {n=5} & {n=1}& {n=2} & {n=3} & $\mathbf{n=4}$ & {n=5}\\
\midrule
Optimal		&3.24	&11.98	&16.16	&15.81	&13.31	&5.32	&13.87	&18.41	&18.93	&15.67\\
almost Equal&3.24	&7.33	&11.6	&13.15	&11.71	&5.32	&13.29	&17.13	&16.97	&14.46\\
Equal		&3.24	&7.06	&8.72	&9.98	&10.37	&5.32	&11.70	&12.52	&13.04	&13.31\\
fixed Equal &3.24	&7.06	&8.72	&9.98	&8.57	&5.32	&11.70	&12.52	&13.04	&10.27 \\
\bottomrule
\toprule
($D, R$) & \multicolumn{5}{c}{(100,250)} & \multicolumn{5}{c}{(100,500)}\\
\cmidrule(r){2-6}\cmidrule(l){7-11}
 & {n=1} & $\mathbf{n=2}$  & {n=3} & {n=4} & {n=5} & {n=1}& {n=2} & {n=3} & $\mathbf{n=4}$ & {n=5}\\
\midrule
Optimal		&7.21	&19.13	&24.90	&24.53	&18.38	&7.84	&22.08	&29.84	&31.11	&26.25\\
almost Equal&7.21	&18.15	&23.17	&21.86	&14.64	&7.84	&20.40	&26.44	&26.18	&19.95\\
Equal		&7.21	&16.42	&15.50	&12.31	&12.42	&7.84	&17.20	&17.66	&18.50	&19.17\\
fixed Equal &7.21	&16.42	&6.79	&6.40	&5.92 	&7.84	&17.20	&17.66	&18.50	&18.42\\
\bottomrule
\toprule
($D, R$) & \multicolumn{5}{c}{(50,350)} & \multicolumn{5}{c}{(60,350)}\\
\cmidrule(r){2-6}\cmidrule(l){7-11}
 & {n=1} & {n=2}  & {n=3} & $\mathbf{n=4}$ & {n=5} & {n=1}& {n=2} & {n=3} & $\mathbf{n=4}$ & {n=5}\\
\midrule
Optimal		&8.76	&17.26	&22.37	&24.08	&22.54	&9.28	&20.98	&27.75	&29.58	&26.69	\\
almost Equal&8.76	&15.78	&19.98	&21.18	&19.51	&9.28	&19.90	&26.35	&27.87	&24.71	\\
Equal 		&8.76	&15.07	&18.67	&18.84	&18.93	&9.28	&18.48	&18.98	&19.11	&19.15	\\
fixed Equal &8.76	&15.07	&18.67	&18.84  &16.32  &9.28	&18.48	&18.98	&19.11	&17.49  \\
\bottomrule
\end{tabular}
\label{table_comparison}
\end{table}

\textit{Discussion :} For a given choice of system parameters, the optimal equal number of MGs for different values of $C$ and $G$ can be precomputed and stored in a look-up table. Then for a specific network scenario (for particular $C$ and $G$), this number is read-off the table and the corresponding total number of combinations of MG subsets can be computed and assigned to available cellular channels efficiently. This offers a practical approach to support multiple multicasts in underlay cellular networks.

\subsection{``Almost'' optimal and computationally efficient solution for the assignment problem}
\label{subsec:assignment}
In this subsection, we compare the performances of MUSCA algorithm, proposed in subsection~\ref{subsec:cas}, and the optimal and exhaustive scheme for the channel assignment problem. For this comparison, for the selection subproblem we use the optimal and exhaustive scheme.

\begin{figure}[!t]
\centering
\includegraphics[width=0.7\hsize]{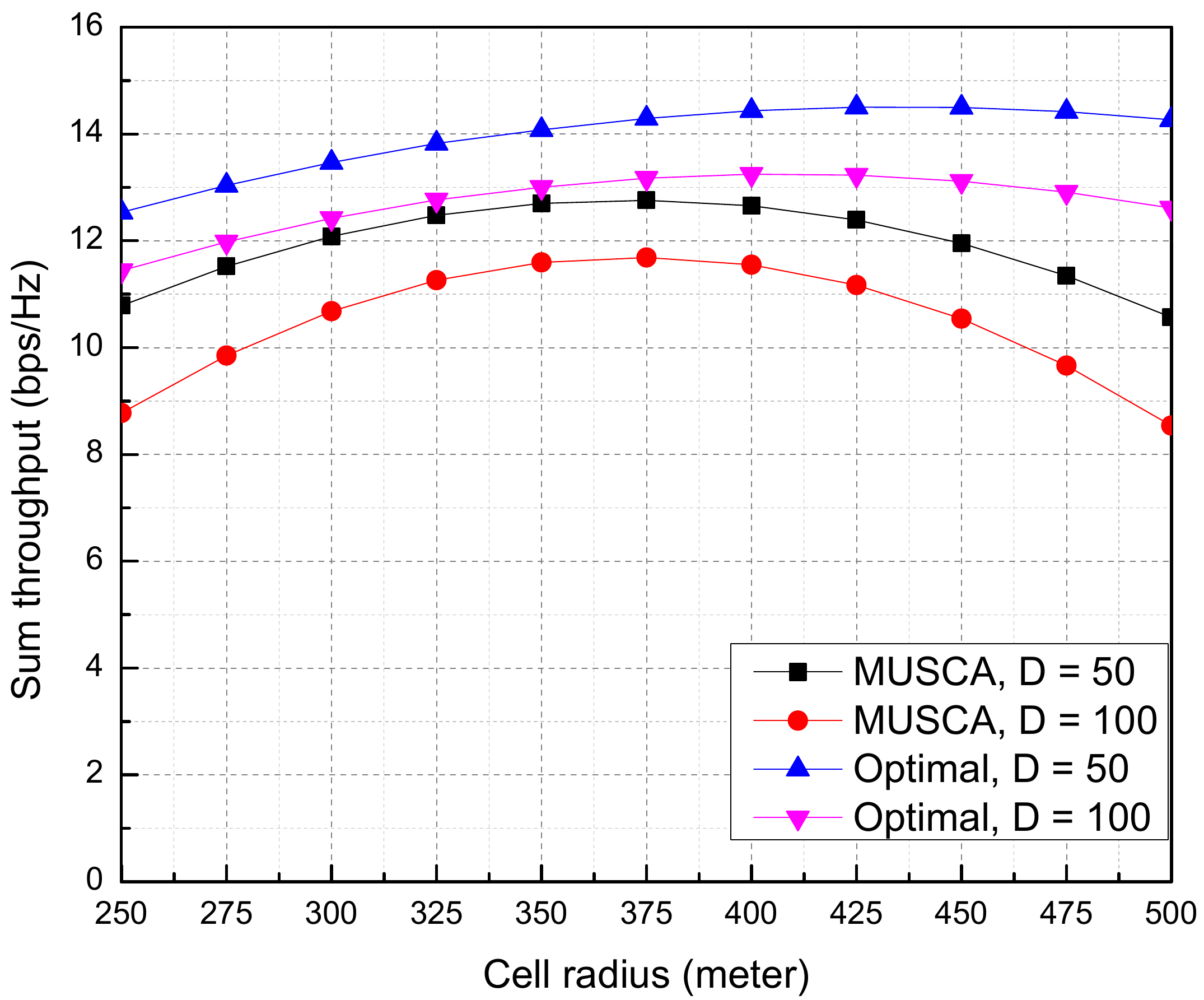}
\caption{The sum throughput as a function of cell radius, for the optimal and MUSCA schemes, $C =3, G=7$.}
\label{fig:c_complxt_w_r}
\end{figure}

Figure~\ref{fig:c_complxt_w_r} depicts the sum throughput as a function of cell radius, $R$, for the optimal scheme and MUSCA algorithm for the channel assignment problem. It can be first observed that the performance of the proposed scheme is close to optimal (loss no more than 1.66 dB), though it is computationally efficient. Also, it can be observed that for both the schemes the sum throughput initially increases with cell radius, then it decreases. This can be explained as follows: for a given value of the exclusion zone radius, $D$, as the cell radius, $R$, initially increases, the co-channel interference among MGs decreases as MGs are more spread out in the cell. However, as $R$ continues to increase, even the receivers of each MG also get more spread out, thus each MGTX has to transmit at higher power, resulting in more co-channel interference among MGs as well to the co-channel CU's uplink.

\begin{figure}[!t]
\centering
\includegraphics[width=0.7\hsize]{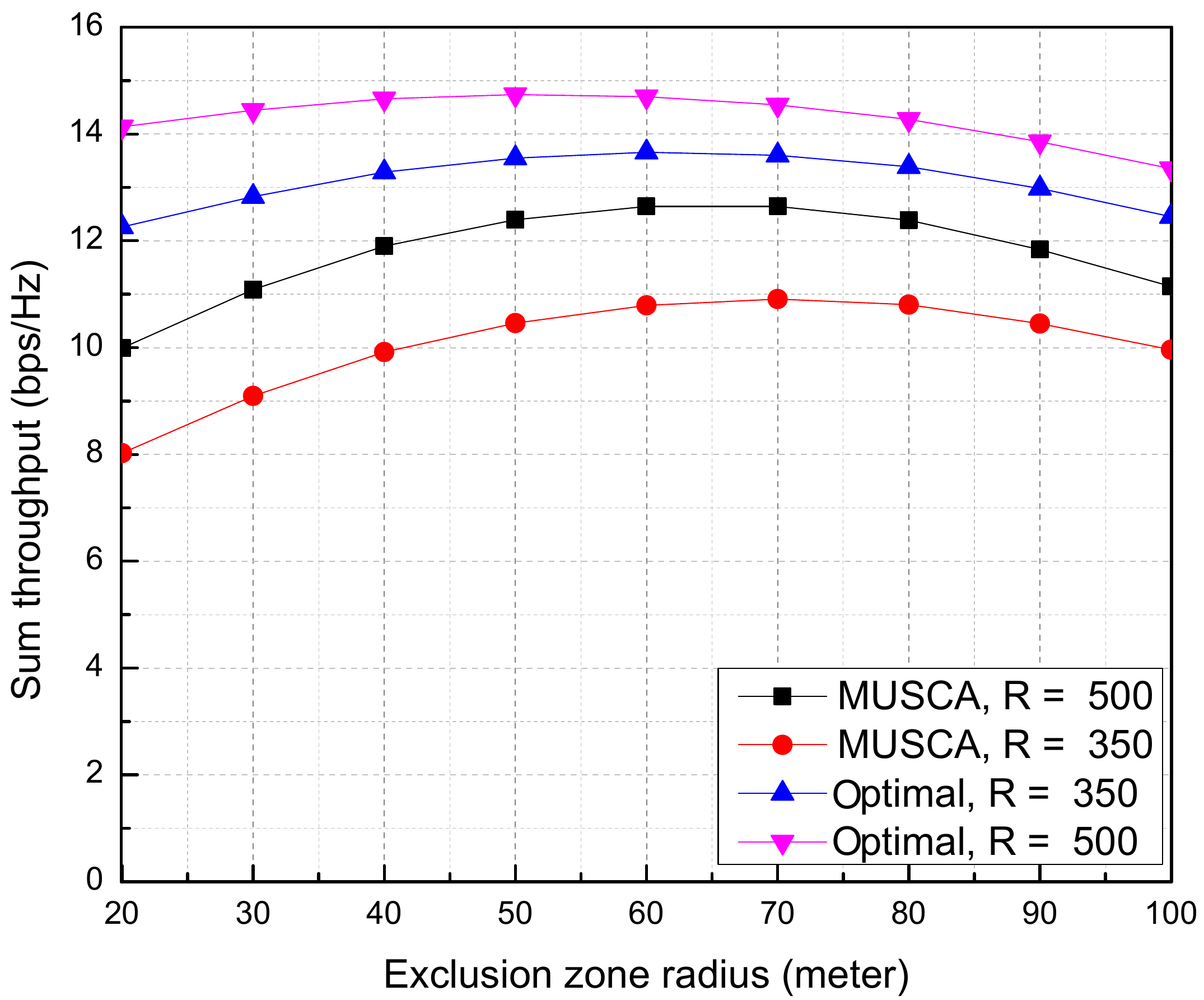}
\caption{The sum throughput as a function of exclusion zone radius, for the optimal and MUSCA schemes, $C =3, G=7$.}
\label{fig:c_complxt_w_d}
\end{figure}

Figure~\ref{fig:c_complxt_w_d} depicts the sum throughput as a function of exclusion zone radius, $D$, for the optimal and computationally efficient scheme for the channel assignment problem. It can be first observed that the performance of the proposed scheme is again close to optimal (loss no more than 1.8 dB), though it is computationally efficient. Also, it can be observed that for both the schemes the sum throughput initially increases with exclusion zone radius, then it decreases. This can be explained as follows: for a given value of the cell radius, $R$, as the exclusion zone radius, $D$, initially increases, the co-channel interference caused by CUs on MGs' receivers reduces, thus leading to increase in the sum throughput. However, as $D$ continues to increase, the co-channel MGs start to get more densely packed, thus increasing mutual interference among them, resulting in lower throughput.

\subsection{Performance evaluation against various system parameters}
\label{subsec:perfeval}
In this subsection, we consider the performance of the computationally efficient fixed-MUSCA algorithm (with $n^*=4$) for the channel allocation problem (a combination of group selection and channel assignment problems) proposed towards the end of Section~\ref{sec:gcas}.

\begin{figure}[!h]
\centering
\includegraphics[width=0.7\hsize]{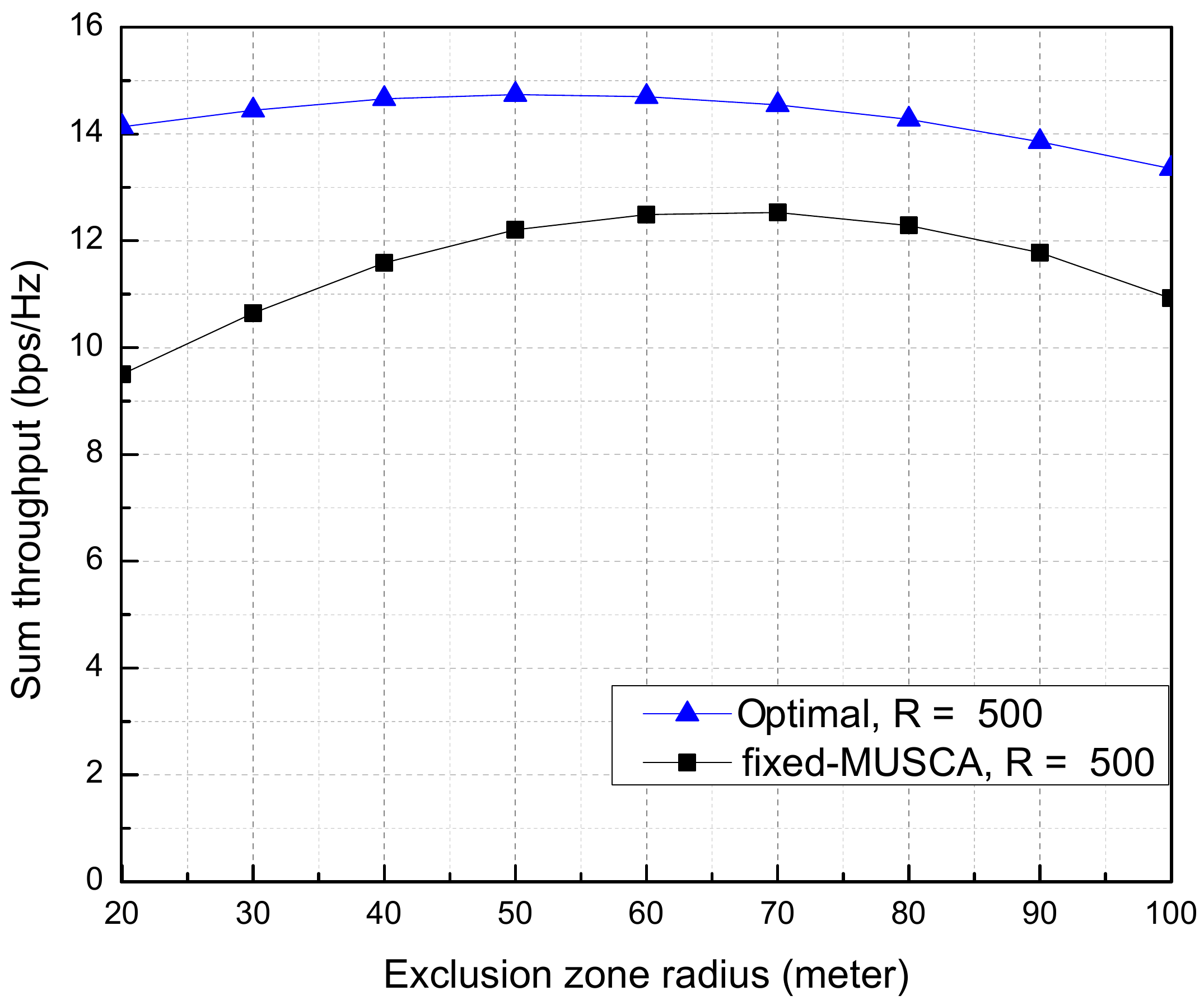}
\caption{The sum throughput as a function of exclusion zone radius, for the optimal and fixed-MUSCA schemes with $G_k = 4$, $\lambda_g = 2e^{-5}$.}
\label{fig:opt_vs_final}
\end{figure}

In Figure~\ref{fig:opt_vs_final} we compare the performance of fixed-MUSCA algorithm with the performance of the optimal and exhaustive scheme. It can be observed that the performance of the proposed scheme is close to the optimal (loss is no more than 1.68 dB), though the computational complexity is exponentially smaller.

\begin{figure}[!t]
\centering
\includegraphics[width=0.7\hsize]{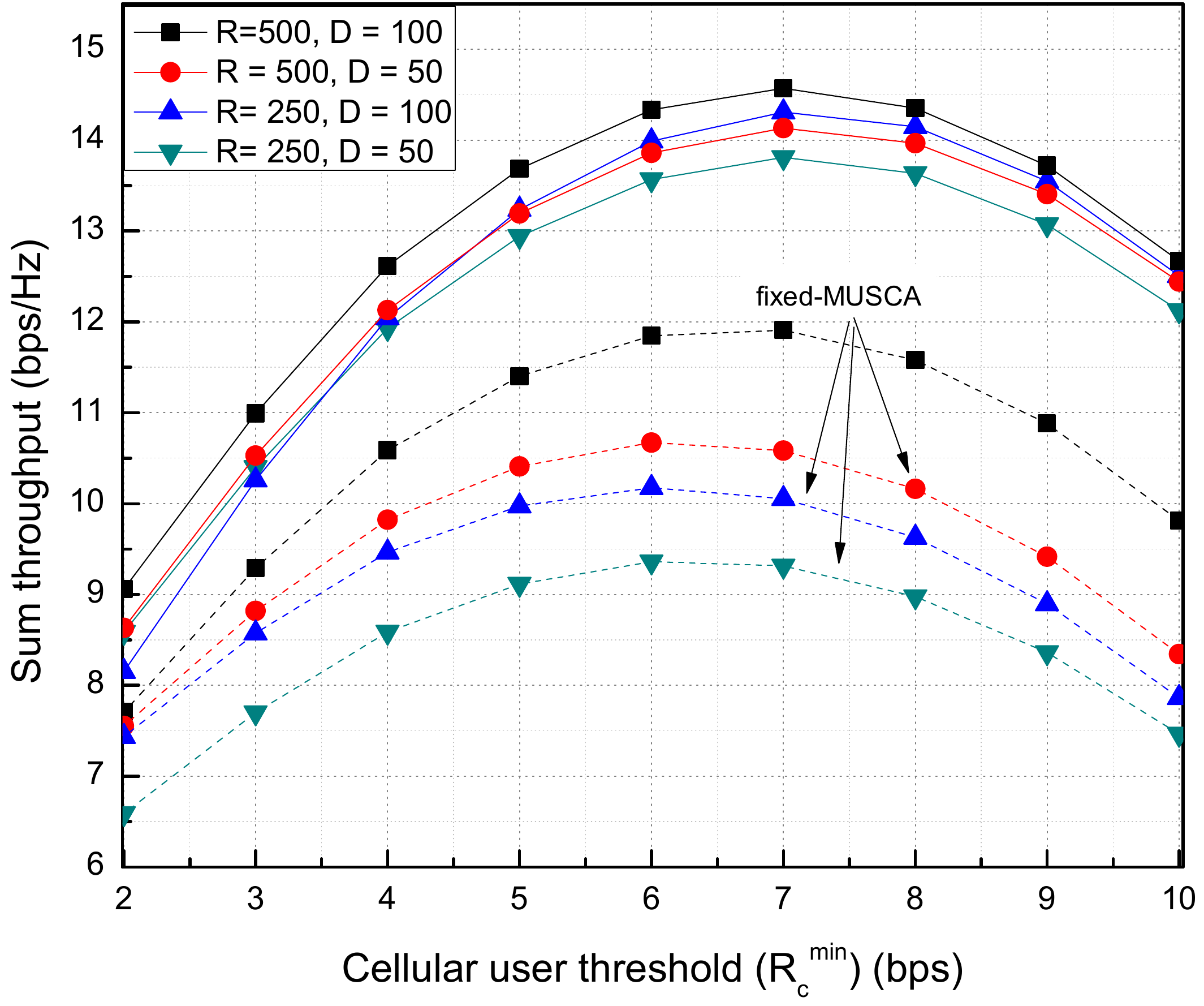}
\caption{The sum throughput as a function of cellular users data rate thresholds, for the optimal and fixed-MUSCA schemes.}
\label{fig:optimal_cu_effi_comparision}
\end{figure}

Figure~\ref{fig:optimal_cu_effi_comparision} depicts the sum throughput obtained by the optimal scheme and fixed-MUSCA algorithm as a function of cellular users thresholds, $R_c^{\min}$, for different values of exclusion zone and cell radii. It can be observed that the sum throughput increases with increase in CU thresholds upto a particular level, then it starts decreasing. This can be explained as follows: when CU threshold is low, the CUs transmit at lower power to fulfill the SIR requirements, therefore, it creates less interference to the MGs receivers. In addition, the MGs may transmit at high power while maintaining the CU thresholds, leading to higher SIR for MGs receivers. However, for high level of cellular users thresholds, the CUs transmit at high power, and cause more interference to MGs receivers, resulting in lower SIR for MGs receivers. It also shows the effectiveness of exclusion zone radius: if exclusion zone radius is large, there are no MG receivers in proximity of CU transmitter, therefore interference is low and the sum throughput is higher. Furthermore, for a large value of the cell radius, the interference among MGs is low, therefore, the sum throughput further increases.

\begin{figure}[!t]
\centering
\includegraphics[width=0.7\hsize]{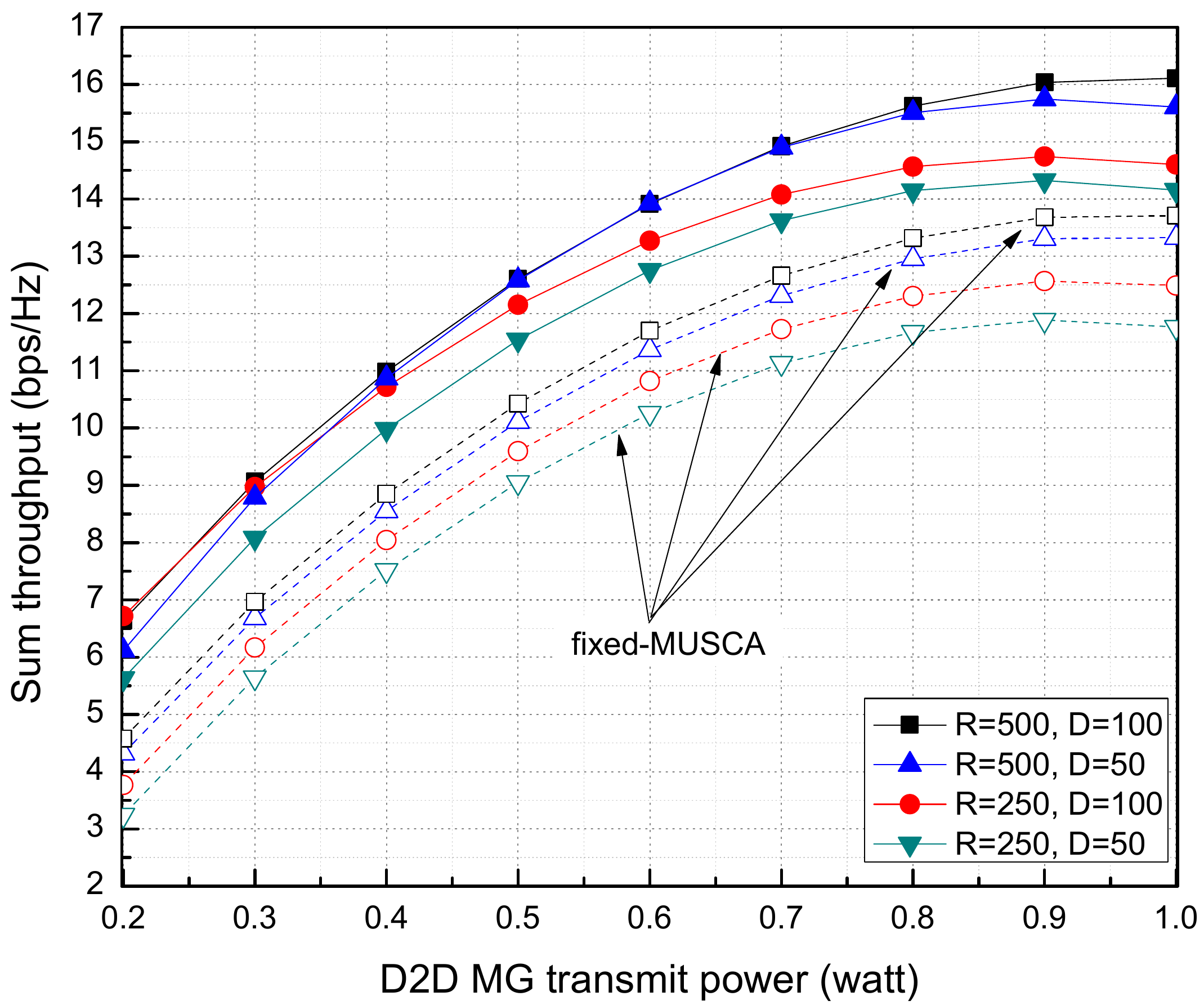}
\caption{The sum throughput as a function of D2D MG transmit power, for the optimal and fixed-MUSCA schemes.}
\label{fig:optimal_cu_effi_power_comparison}
\end{figure}

Figure~\ref{fig:optimal_cu_effi_power_comparison} depicts the behavior of the sum throughput with the limit on maximum D2D MG transmission power, for different exclusion zone and cell radii. It can be observed that the sum throughput increases with increase in transmit power, however, it saturates after some power level. This is because, the maximum power that can be allocated to a MG is $p_{g,k}^{\sup}= \min\{P_G, p_{g,k}^\textrm{high}\}$, and lower values of $P_G$ tightly limit the maximum transmission power, thus the allocated power to MG transmitter is also low. When $P_G$ is high, the supremum is decided by $p_{g,k}^\textrm{high}$, which depends upon other parameters, such as density of MGs, outage thresholds, shown in Lemma \ref{lemma_feasibleRegion}. It can also be observed that, when radius is high, D2D MGs can transmit at the maximum allowable power, without causing much mutual interference. Also, for large value of exclusion zone radius, the interference from CU is low, and SIR of MG receivers is high, thus leading to higher sum throughput.

\begin{figure}[!t]
\centering
\includegraphics[width=0.7\hsize]{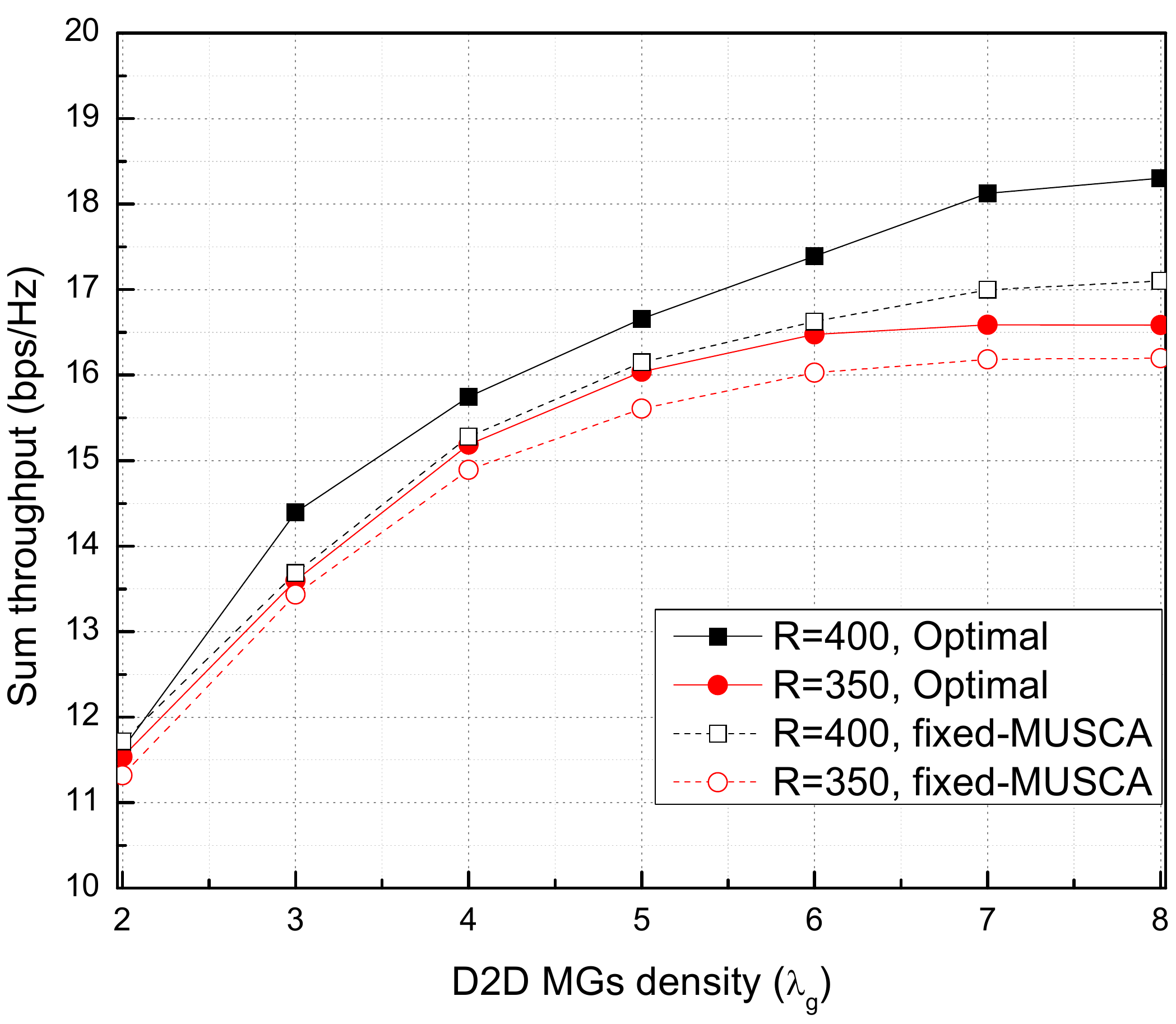}
\caption{The sum throughput as a function of D2D MGs density for different cell radii, for the optimal and fixed-MUSCA schemes.}
\label{fig:density}
\end{figure}

Figure~\ref{fig:density} depicts the sum throughput variation with D2D MGs density, for different cell radii. The sum throughput increases with increase in MGs density, and then it saturates. This can be explained as follows: for lower densities, on average the D2D receivers are closer to the respective MGTX and their SIR thresholds can be met with lower transmit powers without causing excessive interference to neighboring MGs. This leads to almost linear increase in the sum throughput initially. However, as density increases, D2D receivers which are far from MGTX and may also have poor channel conditions invariably start to appear. To meet their SIR constraints MGTXs transmit at higher powers, resulting in higher co-channel interference to other MGs as well as to the corresponding CU's uplink. Eventually the system gets interference saturated and further increase in density does not lead to further increase in the sum throughput.

In all results in this subsection, it can also be observed that the performance of fixed-MUSCA algorithm follows the same trend as the performance of the optimal scheme, with little loss compared to the optimal scheme, but with exponentially lower computational complexity.

\section{Conclusions and Future Work}
\label{sec:concl}
In this work, we propose a pragmatic solution for interference management in underlay D2D multicast communications. Having exclusion zones around cellular users where no receiver of any multicast group can exist, is a realistic approach to effectively reduce the co-channel interference of cellular transmission on D2D multicast reception. Using a stochastic geometry based approach in this scenario, we formulate an optimization problem for sum throughput maximization in D2D-enabled cellular networks, where uplink channels can be reused by multiple multicast groups. We establish that such a problem can be considered as a joint multicast group channel and power allocation problem. The channel allocation problem, itself, can be considered as composed of two subproblems of multicast group subset selection and channel assignment to those subsets. With insights gained from numerical simulations and analysis, we design practical and computationally efficient schemes for these subproblems and thus, for the original channel allocation problem. We also establish \textit{almost} optimal performance of the proposed scheme with respect to variation of various system parameters.

Much work remains to be done. In this work, to simplify some analysis we pessimistically assume that any mobile device which is within the exclusion zone of any cellular user cannot join any multicast group. However, allowing such a device to join a multicast group as long as the target multicast group does not share the channel with a cellular user in whose exclusion zone the said device is, may lead to better performance. Further, we only propose a centralized solution in this work. The design and analysis of a distributed channel and power allocation schemes in this scenario may lead to more practical solutions. Also, we consider only the sum throughput maximization problem in this work. This can be generalized to various utility maximization problems for the service providers and multicast groups. The nature of the optimal solutions in such scenarios may be of independent interest. In the near future, we plan to take up some such problems.

\appendices
\section{Proof of Lemma~\ref{lemma_feasibleRegion}}
\label{appndx:feasible_region}
\setcounter{equation}{0}
\renewcommand{\theequation}{\thesection.\arabic{equation}}
Let $\mathcal{Y}_1 =\gamma_g^\textrm{th} p_{g,k}^{-1} d_{g,r}^\alpha p_{c,k} D^{-\alpha} $, $\mathcal{Y}_2 = \arctan\sqrt{\mathcal{Y}_1}$, $\mathcal{Y}_3= \sqrt{\frac{p_{c,k}}{p_{g,k}}} \sqrt{\gamma_g^{th}} d_{g,r}^{\frac{\alpha}{2}}$, $\mathcal{Y}_4 = \mathcal{Y}_2  + \frac{\sqrt{\mathcal{Y}_1 }}{1+ \mathcal{Y}_1 }$, $\mathcal{Y}_5 = \frac{\gamma_g^{th} d_{g,r}^{\alpha} D^{2 -\alpha}}{ 1 +\mathcal{Y}_1 }$, $ \mathcal{Y}_6 = - \lambda_g^k \frac{\pi^2}{2} \sqrt{\gamma_g^\textrm{th} d_{g,r}^\alpha}$.

From Lemma~\ref{lemma_1}, we have
\begin{align*}
\textrm{Pr}(\gamma_g < \gamma^g_\textrm{th}) &= 1- \mathcal{L}_1 \left( \lambda_c^k, p_{c,k}, D, \gamma_g^\textrm{th} p_{g,k}^{-1} d_{g,r}^\alpha \right) \times \mathcal{L}_0\left( \lambda_g^k, {p_{g,k}}, \gamma_g^\textrm{th} p_{g,k}^{-1} d_{g,r}^\alpha \right)\\
                                             &= 1 - \exp\Bigg(- \lambda_c^k \pi \Bigg\lbrace p_{c,k}^{\frac{1}{2}} \sqrt{\gamma_g^{\textrm{th}}} p_{g,k}^{-\frac{1}{2}} d_{g,r}^{\frac{\alpha}{2}} \Bigg(\bigg( \frac{\pi}{2} -  \arctan\left(\frac{1}{\mathcal{Y}_1}\right)\Bigg) + \frac{\sqrt{\mathcal{Y}_1}}{1+ 
\mathcal{Y}_1}\bigg)\Bigg\rbrace -  \frac{\mathcal{Y}_1 D^2}{ 1 +  \mathcal{Y}_1}\Bigg)\times \\
&\quad\qquad\exp \Bigg( - \lambda_g^k \frac{\pi^2}{2} p_{g,k}^{\frac{1}{2}} \sqrt{\gamma_g^\textrm{th} p_{g,k}^{-1} d_{g,r}^\alpha} \Bigg) \\
                                             &= 1 - \exp \big( - \lambda_c^k \pi \big\lbrace\mathcal{Y}_3 \mathcal{Y}_4  - \mathcal{Y}_5 \big\rbrace +  \mathcal{Y}_6 \big)
\end{align*}

Therefore, we have from \eqref{eq:7}
\begin{align*}
1 - \exp \big( - \lambda_c^k \pi \big\lbrace\mathcal{Y}_3 \mathcal{Y}_4  - \mathcal{Y}_5 \big\rbrace +  \mathcal{Y}_6 \big)  &\leq \Theta_g\\
- \lambda_c^k \pi \big\lbrace \mathcal{Y}_3 \mathcal{Y}_4  - \mathcal{Y}_5 \big\rbrace +  \mathcal{Y}_6 &\geq \ln(1 -  \Theta_g)\\
- \lambda_c^k \pi \Big\lbrace \mathcal{Y}_3 \mathcal{Y}_4  - \mathcal{Y}_5 \Big\rbrace &\stackrel{(a)}{\leq} - \ln(1 -  \Theta_g) - \mathcal{Y}_6\\
\lambda_c^k \pi \sqrt{\frac{p_{c,k}}{p_{g,k}}} \frac{   \sqrt{ 4 \gamma_g^{th} d_{g,r}^{\alpha} p_{c,k} D^{-\alpha}}}{\sqrt{p_{g,k}}}  &\stackrel{(b)}{\leq} - \ln(1 -  \Theta_g) -  \lambda_g^k \frac{\pi^2}{2} \sqrt{\gamma_g^\textrm{th} d_{g,r}^\alpha}  + \gamma_g^{th} d_g^{\alpha} D^{2 - \alpha}  \lambda_c^k \pi \\
p_{g,k} &\geq \frac{\lambda_c^k \pi \sqrt{p_{c,k}}  \sqrt{ 4\gamma_g^{th} d_{g,r}^{\alpha} p_{c,k} D^{-\alpha}}}{- \ln(1 -  \Theta_g) -  \lambda_g^k \frac{\pi^2}{2} \sqrt{\gamma_g^\textrm{th} d_{g,r}^\alpha}  + \gamma_g^{th} d_g^{\alpha} D^{2 - \alpha}\lambda_c^k \pi },
\end{align*}
where $(a)$ follows from the approximation $\arctan(x) = x$, when $x$ is small, and $(b)$ from $1 + \mathcal{Y}_1 \approx 1$ when $\alpha$ is $4$.

Similarly, for an upper bound of feasible region of $p_{g,k}$, we have from Lemma \ref{lemma_2}:
\begin{align*}
\textrm{Pr}(\gamma_c < \gamma_c^{\textrm{th}}) &=  1 - \mathcal{L}_0 \left( \lambda_g^k, p_{g,k}, \gamma_c^\textrm{th} p_{c,k}^{-1} d_{c,b}^\alpha\right) \\ 
&= 1 - \exp \bigg( -\lambda_g^k \frac{\pi^2 \delta}{\sin(\pi \delta)} p_{g,k}^{\delta} \big(\gamma_c^\textrm{th} p_{c,k}^{-1} d_{c,b}^\alpha\big)^\delta \bigg),\\
&=1 - \exp \bigg( - \lambda_g^k \frac{\pi^2}{2} p_{g,k}^{\frac{1}{2}} \sqrt{\gamma_c^\textrm{th} p_{c,k}^{-1} d_{c,b}^4} \bigg), \mbox{ for } \alpha=4
\end{align*}
Again from \eqref{eq:7}, $\textrm{Pr}(\gamma_c < \gamma_c^{\textrm{th}})\leq \Theta_c,$ therefore
\begin{equation*}
1 - \exp \bigg( - \lambda_g^k \frac{\pi^2}{2} p_{g,k}^{\frac{1}{2}} \sqrt{\gamma_c^\textrm{th} p_{c,k}^{-1} d_{c,b}^4} \bigg) \leq \Theta_c \implies p_{g,k} \leq p_{c,k} \Bigg(\frac{ - 2 \ln (1 -  \Theta_c)}{\lambda_g^k {\pi^2}  \sqrt{\gamma_c^{\textrm{th}}} d_{c,b}^2} \Bigg)^2
\end{equation*}

\section{Computational Complexity of Channel Allocation Problem}
\label{appndx:complexity}
Consider the number of possible combinations to be considered for the channel allocation problem, in Equation~\ref{eqn:no_of_cases}. We can provide a lower bound to this number by considering all scenarios where equal number of MGs are assigned to each channel.
\begin{align*}
&\left[\sum_{\substack{g_1 = 1, \ldots, G-(C-1)\\ g_i = 1, \ldots, g_{i-1}, i \in \{2, \ldots, C\}}} \dfrac{{G \choose g_1}{{G-g_1} \choose g_2} \cdots {{G - \sum_{i=1}^{C-1} g_i} \choose g_C}}{\prod_{g=1}^{G-(C-1)} \#_g}\right]C!\\
& \ge \Bigg[\dfrac{{G \choose 1}{{G-1} \choose 1} \cdots {{G - (C-1)} \choose 1}}{C} + \dfrac{{G \choose 2}{{G-2} \choose 2} \cdots {{G - 2(C-1)} \choose 2}}{C} + \ldots + \dfrac{{G \choose \lfloor G/C \rfloor}{{G-\lfloor G/C \rfloor} \choose \lfloor G/C \rfloor} \cdots {{G - (C-1)\lfloor G/C \rfloor} \choose \lfloor G/C \rfloor}}{C}\Bigg]C!\\
& = G!(C-1)!\left[\frac{1}{(1!)^C (G-C)!} + \frac{1}{(2!)^C (G-2C)!} + \ldots + \frac{1}{(\lfloor G/C \rfloor!)^C (G-C\lfloor G/C \rfloor)!}\right]\\
&= O(G!C!).
\end{align*}
The term in the brackets above is smaller than $1$. Note that the terms in square brackets above correspond to the number of combinations for the subset selection subproblem and $C!$ permutations to the channel assignment subproblem.

\section{Examples of computation of combinations for the subset selection problem and its approximations}
\label{appndx:examples}
In this appendix, we discuss examples that illustrate computation of different combinations for the subset selection subproblem and two of its approximations introduced in subsection~\ref{subsec:gss}.

\subsection{Computation of combinations for the subset selection problem}
\label{appndx:subsec:ssp}
Consider a network with $C=3$ and $G=4$. According to Problem P3, we need to consider all possible three non-empty subsets of three to seven MGs as follows. For $q=3$, there is only one possibility of constructing three non-empty subsets, $[1, 1, 1]$. The possibilities for all $q \in \{3, \ldots, 7\}$ are as follows:
\begin{table}[!h]
\centering
\begin{tabular}{ccccc}
$q=3: [1, 1, 1]$, & $q=4: [2, 1, 1]$, & $q=5 : \begin{dcases} [2, 2, 1]\\ [3, 1, 1] \end{dcases}$, & $q=6 : \begin{dcases} [2, 2, 2]\\ [3, 2, 1]\\ [4, 1, 1] \end{dcases}$, & $q=7 : \begin{dcases} [3, 2, 2], [3, 3, 1]\\ [4, 2, 1], [5, 1, 1] \end{dcases}$\\
\end{tabular} 
\end{table}

Therefore, the number of combinations of MG subsets to consider for each value of $q$ are as follows:

\begin{align*}
q=3 &: \frac{1}{3}{7 \choose 1}{6 \choose 1}{5 \choose 1} = 70\\
q=4 &: \halfrac{7 \choose 2}{5 \choose 1}{4 \choose 1} = 210\\
q=5 &: \halfrac{7 \choose 2}{5 \choose 2}{3 \choose 1} + \halfrac{7 \choose 3}{4 \choose 1}{3 \choose 1}= 525\\
q=6 &: \frac{1}{3}{7 \choose 2}{5 \choose 2}{3 \choose 2} + {7 \choose 3}{4 \choose 2}{2 \choose 1} + \halfrac{7 \choose 4}{3 \choose 1}{2 \choose 1} = 735\\
q=7 &: \halfrac{7 \choose 3}{4 \choose 2}{2 \choose 2} + \halfrac{7 \choose 3}{4 \choose 3}{1 \choose 1} + {7 \choose 4}{3 \choose 2}{1 \choose 1} + \halfrac{7 \choose 5}{2 \choose 1}{1 \choose 1} = 301
\end{align*}

These $1841$ combinations of subsets of MGs are the solution of the subset selection subproblem.

Further each of these combination of subsets is given as the input to the channel assignment subproblem, which generates all $C! = 3!$ permutations of each of these subset combinations over three channels. If the three channels in this problem are labeled $\{1, 2, 3\}$ and the seven MGs in this example are labeled $\{a, b, c, d, e, f, g\}$, then for $q=3$ one of the combinations of MG subsets is $\{\{a\},\{b\},\{c\}\}$. These three subsets can be assigned to three channels in six ($3!$) ways, namely
\begin{table}[!h]
\centering
\begin{tabular}{c|c|c|c|c|c}
$1: \{a\}$ & $1: \{a\}$ & $1: \{b\}$ & $1: \{b\}$ & $1: \{c\}$ & $1: \{c\}$\\
$2: \{b\}$ & $2: \{c\}$ & $2: \{a\}$ & $2: \{c\}$ & $2: \{a\}$ & $2: \{b\}$\\
$3: \{c\}$ & $3: \{b\}$ & $3: \{c\}$ & $3: \{a\}$ & $3: \{b\}$ & $3: \{a\}$
\end{tabular} 
\end{table}

In the same way, we generate six permutations over three channels for other $69$ combinations of MG subsets for $q=3$. Then, all these permutations together form the set $\Pi_3^3$ in Problem P3. Further, this way we generate the sets $\Pi_q^3$ for the other values of $q$. All these sets together form the solution of the channel assignment subproblem.

\subsection{Computation of combinations for the almost equal subset selection problem}
\label{appndx:subsec:aessp}
Consider the network in the previous subsection with $C=3$ and $G=4$. According to Problem P4, we need to consider all possible \textit{almost} equal three non-empty subsets of three to seven MGs as follows. For $q=3$, there is only one possibility of constructing three non-empty subsets, $[1, 1, 1]$. The possibilities for all $q \in \{3, \ldots, 7\}$ are as follows:
\begin{table}[!h]
\centering
\begin{tabular}{ccccc}
$q=3: [1, 1, 1]$, & $q=4: [2, 1, 1]$, & $q=5 : [2, 2, 1]$, & $q=6 : [2, 2, 2]$, & $q=7 : [3, 2, 2]$\\
\end{tabular} 
\end{table}

Therefore, the number of combinations of MG subsets to consider for each value of $q$ are as follows:

\begin{align*}
q=3 &: \frac{1}{3}{7 \choose 1}{6 \choose 1}{5 \choose 1} = 70\\
q=4 &: \halfrac{7 \choose 2}{5 \choose 1}{4 \choose 1} = 210\\
q=5 &: \halfrac{7 \choose 2}{5 \choose 2}{3 \choose 1} = 315\\
q=6 &: \frac{1}{3}{7 \choose 2}{5 \choose 2}{3 \choose 2} = 210\\
q=7 &: \halfrac{7 \choose 3}{4 \choose 2}{2 \choose 2} = 70
\end{align*}

These $875$ combinations of subsets of MGs are the solution of the subset selection subproblem P4, which are significantly less than $1841$ solutions for the subset selection problem in P3.

\subsection{Computation of combinations for the equal subset selection problem}
\label{appndx:subsec:essp}
Consider again the network in the previous subsection with $C=3$ and $G=4$. According to Problem P5, we need to consider all possible equal three non-empty subsets of three to seven MGs as follows. For $q=3$, there is only one possibility of constructing three non-empty subsets, $[1, 1, 1]$. The possibilities for all $q \in \{3, \ldots, 7\}$ are as follows:
\begin{table}[!h]
\centering
\begin{tabular}{cc}
$q=3: [1, 1, 1]$, & $q=6 : [2, 2, 2]$\\
\end{tabular} 
\end{table}

Therefore, the number of combinations of MG subsets to consider for each value of $q$ are as follows:

\begin{align*}
q=3 &: \frac{1}{3}{7 \choose 1}{6 \choose 1}{5 \choose 1} = 70\\
q=6 &: \frac{1}{3}{7 \choose 2}{5 \choose 2}{3 \choose 2} = 210
\end{align*}

These $280$ combinations of subsets of MGs are the solution of the subset selection subproblem P5, which are significantly less than $875$ solutions for the subset selection problem in P3.

{
\singlespacing

}
\end{document}